\documentclass[onecolumn]{aastex631}

\shorttitle{Extended MCDHF calculations of energy levels and transition data for \ion{N}{1}}
\shortauthors{Li et al.}
\usepackage{multirow}
\usepackage{threeparttablex}
\usepackage{tablefootnote}
\usepackage{amsmath}	% Advanced maths commands
\usepackage{amssymb}	% Extra maths symbols
\usepackage{ulem}

\graphicspath{{./}{figures/}}
\newcommand{\lgeps}{A\left(\mathrm{N}\right)}
\newcommand{\lggf}{\log(gf)}

\begin{document}
%\linenumbers
\title{Extended MCDHF calculations of energy levels and transition data for \ion{N}{1}}

\author{M.~C.~Li}
\affiliation{School of Electronic Information and Electrical Engineering, Huizhou University, Huizhou, China 516007}

\author{W.~Li}
\affiliation{Key Laboratory of Solar Activity, National Astronomical Observatories, Chinese Academy of Sciences, Beijing 100012, China}
\correspondingauthor{W. Li}
\email{wxli@nao.cas.cn}

\author{P.~J\"onsson}
\affiliation{Department of Materials Science and Applied Mathematics, Malm\"o University, SE-205 06 Malm\"o, Sweden}

\author{A.~M.~Amarsi}
\affiliation{Theoretical Astrophysics, Department of Physics and Astronomy, Uppsala University, Box 516, SE-751 20 Uppsala, Sweden}

\author{J.~Grumer}
\affiliation{Theoretical Astrophysics, Department of Physics and Astronomy, Uppsala University, Box 516, SE-751 20 Uppsala, Sweden}

\begin{abstract}
Accurate and extensive atomic data are essential for spectroscopic analyses of stellar atmospheres and other astronomical objects. We present energy levels, lifetimes, and transition probabilities for neutral nitrogen, the sixth most abundant element in the cosmos. The calculations employ the fully relativistic multiconfiguration Dirac-Hartree-Fock and relativistic configuration interaction methods, and span the 103 lowest states up to and including $\mathrm{2s^22p^25s}$. Our theoretical energies are in excellent agreement with the experimental data, with an average relative difference of 0.07\%. In addition, our transition probabilities are in good agreement with available experimental and theoretical data. We further verify the agreement of our data with experimental results via a re-analysis of the solar nitrogen abundance, with the results from the Babushkin and Coulomb gauges consistent to 2\% or 0.01 dex. {We estimated the uncertainties of the computed transition data based on a statistical analysis of the differences between the transition rates in Babushkin and Coulomb gauges. Out of the 1701 computed electric dipole transitions in this work, 83 (536) are associated with uncertainties less than 5\% (10\%).} 

\end{abstract}

\keywords{atomic data; solar abundances}

\section{Introduction} \label{sec:intro}
Nitrogen is the sixth most abundant element in the Universe~\citep{asplund.2021}, 
and its abundance is an important diagnostic in the study of
the structure and evolution of stars \citep{2007A&A...461..571H,2014ApJ...781...88A,2014A&A...565A..39M},
globular clusters \citep{2022arXiv220910219S},
and galaxies \citep{2017MNRAS.469..151B,2015MNRAS.453.1855M,2017MNRAS.465..501S,vincenzo.2018,esteban.2020}.
%the study of the chemical evolution of galaxies~\citep{vincenzo.2018,esteban.2020}. \sout{The determination of nitrogen abundances in stellar atmospheres are strongly dependent upon \sout{the} high quality atomic data, \pj{such as} \sout{for example} weighted oscillator strengths and transition probabilities.} 
%\pj{High quality atomic data for \ion{N}{1} are indispensable for analyses of spectra observed from stellar atmospheres.} 
Nitrogen abundance measurements in the atmospheres of stars via stellar spectroscopy {are} critical
in this endeavour. In hot stars of spectral types O, B, A, and F,
near-infrared and infrared \ion{N}{1} lines arising from transitions 
involving the configurations 
$\mathrm{2p^2(^3P)3s}$, $\mathrm{2p^2(^3P)3p}$, $\mathrm{2p^2(^1D)3s}$, $\mathrm{2p^2(^3P)3d}$, and $\mathrm{2p^2(^1D)3p}$
are typically used \citep{takeda.1995,przybilla.2001}.
Several weak \ion{N}{1} lines can also be detected in 
cooler stars \citep{2022AJ....164...87K}, 
including the Sun; these have been used to inform the solar nitrogen abundance,
together with molecular diagnostics such as NH and CN
\citep{1978MNRAS.182..249L,1990A&A...232..225G,2020A&A...636A.120A,2021A&A...656A.113A}.
%\ama{[Suggest to use ``mathrm'' for configurations? \jon{Maybe - this is different for different journals - LS terms I think are always rm, while configurations sometimes are italics. Check the journal formatting guidelines.} currently it is quite hard for the reader, with many italicised letters and numbers... One might also add dots "." between the orbitals e.g. $\mathrm{2p^2.3s}$... \jon{no, this is sometimes ok for ascii data tables, but in formatted text there should be no dots} Or at least, avoid having multiple "configuration dash configuration", like ${2p^2(^3P)3p}$ -- ${2p^2(^3P)3p}$, appearing next to each other. \jon{yes I agree with this, better to make a list then.}]}

%For example, near-infrared and infrared spectral lines arising from transitions between $2p^2(^1D)3s-2p^2(^1D)3p$, $2p^2(^3P)3p-2p^2(^3P)3d$  and $2p^2(^3P)3s-2p^2(^3P)3p$  configurations are observed in Late-B, F supergiants and BA-type stars~\citep{takeda.1995,przybilla.2001}. These spectral lines can be used in nitrogen abundances and stellar parameters analyses~\citep{przybilla.2001,amarsi.2020,andrievsky.2021}.

%due to the different time scales it is ejected to the interstellar medium.The energy levels of \ion{N}{1} presented in NIST Atomic Levels and Spectra Database \citep{NIST_ASD.2021}

High quality atomic data for \ion{N}{1} are indispensable for reliable stellar spectroscopic analyses.
During the past 30 years, {several tens of} experimental and theoretical studies of transition data have been carried out for \ion{N}{1}. {The complete lists of published papers on these measurements and calculations can be retrieved from the NIST Atomic Transition Probability Bibliographic database~\citep{kramida.2010}}. {For example,} on the experimental side, using a wall-stabilized arc source, measurements of transition probabilities and line strengths for visible and infrared lines, originating from transition arrays 3s-3p, 3s-4p, 3p-3d, and 4d-5s
%$2p^2(^1D)3s-2p^2(^1D)3p$, $2p^2(^3P)3s-2p^2(^3P)3p$, $2p^2(^3P)3s-2p^2(^3P)4p$, $2p^2(^3P)3p-2p^2(^3P)3d$, $2p^2(^3P)3p-2p^2(^3P)4d$, and $2p^2(^3P)3p-2p^2(^3P)5s$
%\jon{ \begin{itemize}
%    \item[] $2p^2(^1D)3s - 2p^2(^1D)3p$
%    \item[] $2p^2(^3P)3s - 2p^2(^3P)3p$, $2p^2(^3P)4p$
%    \item[] $2p^2(^3P)3p - 2p^2(^3P)3d$, $2p^2(^3P)4d$, $2p^2(^3P)5s$
%\end{itemize} }
were performed by~\citet{musielok.1995}, \cite{baclawski.2002},~\citet{Baclawski.2010,baclawski.2008}, and~\citet{bridges.2010}. The relative uncertainties of these measurements were claimed to be no more than 15\%. 
Using the same method, \citet{goldbach.1991} and~\citet{goldbach.1992} measured the oscillator strengths for 19 lines in the 90--125 nm region and 18 lines in the 120--200 nm region, respectively.
There are also a number of experimental measurements of
lifetimes and oscillator strengths using various techniques~\citep{bengtsson.1992,dumont.1974,catherinot.1979,copeland.1987,Bromander.1978}.

On the theoretical side, {a few tens of} theoretical studies of excitation energies and
transition data for \ion{N}{1} have been reported. For example,~\citet{hibbert.1991} calculated the
excitation energies and oscillator strengths for a number of dipole-allowed and intercombination
transitions between doublet and quartet states with the configuration interaction method.
Using the multiconfigurational Hartree-Fock and
Breit-Pauli (MCHF-BP) method, calculations of oscillator strengths for \ion{N}{1} have been
performed by~\citet{tong.1994}, 
\citet{tachiev.2002} and~\citet{fischer.2004}. \citet{tong.1994} only reported oscillator strengths of transitions among low-lying quartet states, whereas 
\citet{tachiev.2002} and \citet{fischer.2004} calculated energy levels and lifetimes for all levels up to $\mathrm{2s^22p^23d}$, as well as transition data between these levels. Among these theoretical results, the MCHF-BP values~\citep{tachiev.2002,fischer.2004} are in overall better agreement with experimental results than the others. 
More recently, \cite{2022arXiv220614095B} calculated the $gf$-values of the two lines at 8683 \AA~and 8629 \AA,
which are diagnostics of the solar nitrogen abundance, {using} a combination of different methods, i.e. AUTOSTRUCTURE based on the Thomas-Fermi-Dirac-Amaldi central potential,
pseudo-relativistic Hartree-Fock, and multiconfiguration Dirac-Hartree-Fock (MCDHF) methods.

In the present work we perform large-scale ab initio calculations of excitation energies and electric dipole (E1) transition parameters (transition rates, line strengths and weighted oscillator strengths) for the 103 lowest states belonging to the $\mathrm{2s^22p^3,~2s2p^4,~2s^22p^2{\textit{nl}}~(\textit{n}=3,4,~\textit{l}=s,p,d,f)}$, and $\mathrm{2s^22p^25s}$ configurations in \ion{N}{1}. Calculations are based on fully relativistic MCDHF and configuration interaction (RCI) methods, as implemented in the general-purpose relativistic atomic structure package GRASP2018\footnote{See also the open-source GitHub repository maintained by the CompAS collaboration: \url{https://github.com/compas/grasp}.}~\citep{fischer.2019}.

\section{Theory and Computations}

\subsection{Theory}\label{theory}
In the MCDHF method \citep{fischer.2016} as implemented in the GRASP code, the atomic eigenstate is represented by an atomic state expressed as a linear combination of configuration state functions (CSFs) with equal parity $P$ and angular momentum quantum numbers $JM$:
\begin{equation}\label{ASFs}
\centering
\Psi(\gamma PJM )= \sum_{i} c_{i}\Phi(\gamma_{i} PJM )
\end{equation}
The CSFs are $jj$-coupled many-electron functions built from antisymmetrized products of single-electron Dirac orbitals. The quantities $c_i$ and $\gamma_{i}$ are, respectively, the mixing coefficient and additional labeling needed to uniquely specify each CSF. 
The radial parts of the Dirac orbitals and the expansion coefficients $c_i$ of the targeted states are all optimized to self-consistency by solving the MCDHF equations, which are derived by applying the variational principle on the weighted {average energy} of the targeted states.
Higher-order electron-electron interactions, such as the frequency-independent Breit interaction and leading quantum electrodynamical effects in the form of the self-energy (SE) and the screened vacuum polarization (VP), are included in the subsequent relativistic configuration interaction (RCI) calculations using the orbital basis from the MCDHF optimization~\citep{grant.2007}. 

The E1 transition data (transition probabilities and weighted oscillator strengths) between two states $\gamma 'P'J'$ and $\gamma PJ$ are expressed in terms of reduced matrix elements of the electric dipole transition operator $\rm{T}^{(1)}$. From Equation~\ref{ASFs}, these matrix elements can be written as
\begin{equation} \label{T}
    \langle \Psi(\gamma PJ)\| {\rm T}^{(1)}\| \Psi(\gamma'P'J') \rangle = 
     \sum_{i,j} c_{i}c_{j}'\langle \Phi(\gamma_{i} PJ)\|{\rm T}^{(1)}\|\Phi(\gamma_{j}' P'J' ) \rangle.
\end{equation}
%\jon{Comment: the labelling is a bit confusing: for ASF's we tend to use captital gamma $\Gamma$ for the additional labels, and then I'm not sure if it's a good idea to use an the apostrophe on the $c_j$ and $\gamma_j$? Since they are already labelled with the j index?}
Here, $ c_{i}$ and $c_{j}$ are, respectively, the expansion coefficients of the CSFs for the upper and lower states.
Using the Brink-and-Satchler convention, the reduced matrix elements in Equation~\ref{T} can be expressed in terms of spin-angular coefficients $d_{ab}^{(1)}$ and operator strengths as
\begin{equation}
\langle\Phi(\gamma_{i} PJ )\| {\rm T}^{(1)}\|\Phi(\gamma_{j}' P'J' )\rangle = 
 \sum_{a,b} d_{ab}^{(1)}\langle n_al_aj_a\|{\rm t}^{(1)}\|n_bl_bj_b\rangle
\end{equation}
where
\begin{eqnarray}
\langle n_al_aj_a\|{\rm t}^{(1)}\|n_bl_bj_b\rangle=  
\left(
\frac{(2j_b+1)\omega}{\pi c}
\right)^{1/2} (-1)^{j_a-1/2}
\left(
\begin{array}{ccc}
j_a & 1 & j_b \\
\frac{1}{2} & 0 & -\frac{1}{2}
\end{array}
\right)\overline{M_{ab}}
\end{eqnarray}

Here, {$n, l, j$ are, respectively, {the} principal, orbital and angular quantum numbers {of spin-orbitals a and b}, {$\hbar \omega$ is the electromagnetic energy}, and $\overline{M_{ab}}$ is the radiative transition integral defined by~\citet{grant.1974}. For electric type multipoles, the $\overline{M_{ab}}$ integral can be written as $\overline{M_{ab}}(G)=\overline{M_{ab}^e}+G\overline{M_{ab}^l}$ (see equation 4.10 in \citet{grant.1974}), where $\overline{M_{ab}^e}$ is the Coulomb gauge integral , $\overline{M_{ab}^l}$ the longitudinal part} and $G$ the gauge parameter. Two familiar choices, corresponding to the velocity and length operator forms in non-relativistic theory, are the Coulomb ($G=0$) and Babushkin ($G = \sqrt{2}$) gauges.

The agreement between transition rates in the Coulomb gauge ($A_\mathrm{C}$) and the Babushkin gauge ($A_\mathrm{B}$), {is} often taken as {an} internal indicator of accuracy of calculated data, especially when there are no experimental results available. %\ama{a lower bound on the uncertainty of the computed rates [fair to call it a lower bound? \jon{Jon: No, the agreement/disagreement can be accidental so it's not bounding anything as far as I can understand. It has to be combined with a convergence analysis to say something meaningful.}]}
{The relative differences between transition rates $A_\mathrm{B}$ and $A_\mathrm{C}$, d$\it{T}$, is defined as \citep{fischer.2009, ekman.2014}}
\begin{equation}\label{dT}
\rm{d}\it{T}=\frac{|A_\mathrm{B}-A_\mathrm{C}|}{\max(A_\mathrm{B}, A_\mathrm{C})}.
\end{equation}
{It should be emphasized that $\rm{d}\it{T}$ gives an estimation of the uncertainty for groups of lines in a statistical manner.}

\subsection{Computational Schemes}
Calculations were performed in the extended optimal level (EOL)
scheme \citep{dyall.1989} for the weighted average of the even and odd parity states.
These states belong to the \{$\mathrm{2s2p^4}$, $\mathrm{2s^22p^2}$$nl(n=3,4, l=\mathrm{s,d)}$, $\mathrm{2s^22p^25s}$\} even configurations 
and the \{$\mathrm{2s^22p^3}$, $\mathrm{2s^22p^2}nl(n=3,4, l=\mathrm{p,f)}$\} odd configurations. These target configurations are included in the multireference (MR) 
set used in the MCDHF calculations. 

Following the CSF generation strategies used by, e.g. \citet{papoulia.2019.v7} and~\citet{li.2021.v502},
the MCDHF calculations were based on CSF expansions for which we impose restrictions on the orbital excitations from the deeper subshells. The CSFs are generated and systematically enlarged through single and double (SD) excitation from subshells occupied in the predefined MR configurations to an active set (AS) of orbitals in a step-by-step manner \citep{olsen.1988, sturesson.2007, fischer.2016}. 

In the present calculations, the orbitals in the $n=1$ shell of the MR configurations are defined as core orbitals.
The remaining orbitals are defined as valence orbitals. Based on these definitions, the valence-valence (VV) electron correlations are included in the calculation by allowing SD excitations from the valence orbitals to active sets of orbitals, with the restriction that there is at most one excitation from the $n=2$ shell.

With an MCDHF orbital basis at hand, the MR set was then further extended and applied in a subsequent RCI calculation, as shown in Table~\ref{tab:schemes}. {The extended MR set comprises configurations that give rise to $LSJ$-coupled CSFs with weights larger than 0.05.} Core-valence (CV) electron correlation effects were taken into account by allowing SD substitutions from the valence shells and the $\mathrm{1s^2}$ core of the configurations in the extended MR, with the restriction that at most one excitation is allowed from $\mathrm{1s^2}$, to the final active set, \{$\mathrm{11s, 10p, 10d, 10f, 7g, 6h}$\}.
The numbers of CSFs in the final even and odd state expansions were, respectively, 13~431~751 and 17~662~086, distributed over the different $J$ symmetries.
\begin{table*}[h!]
\setlength{\tabcolsep}{8mm}
\caption{\label{tab:schemes}Summary of the active space construction. The configurations in the second and third columns, respectively, represent the target states and the extended target MR set in RCI calculation. $N_\mathrm{CSFs}$ denotes the number of CSFs in the total expansion for the even and odd parity states in the final RCI calculation.}
\centering
\begin{tabular}{cccc}
\hline\hline
Parity & MR & Extended MR in RCI & $N_\mathrm{CSFs}$ \\
\hline 
& $\mathrm{2s2p^4, 2s^22p^23s}$ & $\mathrm{2s2p^3}$$n$p($3\leq n\leq5$) &  \\
Even & $\mathrm{2s^22p^23d, 2s^22p^24s}$& $\mathrm{2p^4}$$nl$($3\leq n\leq5,l=\mathrm{s,d}$)
& 13 431 751 \\
& $\mathrm{2s^22p^24d, 2s^22p^25s}$ &  & \\
\hline
& $\mathrm{2s^22p^3, 2s^22p^23p}$ &  $\mathrm{ 2s2p^3}$$nl$($3\leq n\leq5,l=\mathrm{s,d}$)& \\
Odd & $\mathrm{2s^22p^24p, 2s^22p^24f}$ &$\mathrm{2p^44f}$, $\mathrm{ 2s2p^4}$$n$p($3\leq n\leq5$) & 17 662 086 \\
\hline
\end{tabular}
\end{table*}

\section{Results and Discussion}
\subsection{Energies and Lifetimes}
The calculated energies and corresponding wave function composition in $LS$-coupling for the lowest 103 states are displayed in Table~\ref{tab:energy}. 
The labelling of the eigenstates is defined by the {$LSJ$}-coupled CSF with the largest expansion coefficient resulting from the transformation from $jj$-coupling to $LSJ$-coupling using the method by~\citet{Gaigalas.2017.V5.p6}. {We note that $LSJ$-coupling might not be the best {representation} to describe the {states associated to the} $\mathrm{2s^22p^24f}$ configuration and that the National Institute of Standards and Technology Atomic Spectra Database (NIST-ASD)~\citep{NIST_ASD.2021} adopts $JK$-coupling {in their labeling of these}.} 
The corresponding experimental energies provided via NIST-ASD, as well as radiative lifetimes in both Babushkin (B) and Coulomb (C) forms are also included in Table~\ref{tab:energy}. 
{When compared with energies from NIST-ASD, we noticed two pairs of levels, \#74-75 (2p$^2$4d~$^4$P$_{5/2}$ and $^2$F$_{5/2}$) and \#80-82 (2p$^2$4d~$^4$D$_{1/2}$ and $^4$P$_{1/2}$), being inverted based on the energy ordering. 
A closer inspection of the $LS$-composition reveals that these states are strongly mixed, the leading percentages being about 40-60\% and the second components about 30-40\%, and they might be less accurately described by the conventional labelling based on the expansion coefficients. 
Since a close agreement of the computed energies with the NIST-ASD values being observed for the other levels, we matched the corresponding two pairs of levels with experimental energies based on the energy ordering within each symmetry. Note that the labelling of these two pairs of levels, i.e. \#74-75 and \#80-82, given in the second column of Table \ref{tab:energy} are inverted compared to that of the NIST-ASD data.
}
\begin{table*}[h!]
\setlength{\tabcolsep}{4mm}
\caption{\label{tab:energy}Energies (in cm$^{-1}$) and lifetimes (in s) in both the Babushkin ($\tau_B$) and Coulomb ($\tau_C$) gauges for the lowest 103 levels of \ion{N}{1}. Energy levels are given relative to the ground state and compared with results from NIST-ASD \citep{NIST_ASD.2021}. {The labelling in the second column is defined by the composition with the largest expansion coefficient given in the third column.} This table is available in its entirety in electronic form.}
\centering
\begin{tabular}{lllrrrcc}
\hline\hline
No. & State & $LS$-composition & $E_{\text{RCI}} $ & $E_{\text{NIST}} $ & \colhead{$\Delta E$} &$\tau_B$ & $\tau_C$ \\
\hline
1   & $\mathrm{2s^{2}\,2p^{3}~^{4}S_{3/2}^{\circ}}$            & 0.91                                                                                                                                               & 0           &    0.000       &  0  &             &            \\       
2   & $\mathrm{2s^{2}\,2p^{3}~^{2}D_{5/2}^{\circ}}$            & 0.89                                                                                                                                               & 19426       &    19224.464   &  -201   &             &            \\    
3   & $\mathrm{2s^{2}\,2p^{3}~^{2}D_{3/2}^{\circ}}$            & 0.89                                                                                                                                               & 19435       &    19233.177   &  -201   &             &            \\    
4   & $\mathrm{2s^{2}\,2p^{3}~^{2}P_{1/2}^{\circ}}$            & 0.85 + 0.03~$\mathrm{2p^{5}~^{2}P^{\circ}}$                                                                                                        & 29142       &    28838.920   &  -304   &             &            \\    
5   & $\mathrm{2s^{2}\,2p^{3}~^{2}P_{3/2}^{\circ}}$            & 0.85 + 0.03~$\mathrm{2p^{5}~^{2}P^{\circ}}$                                                                                                        & 29143       &    28839.306   &  -303   &             &            \\    
6   & $\mathrm{2s^{2}\,2p^{2}(^{3}P)\,3s~^{4}P_{1/2}}$          & 0.82 + 0.12~$\mathrm{2s\,2p^{4}~^{4}P}$                                                                                           & 83332       &    83284.070    &  -48    &  2.44E-09 &  2.44E-09\\                  
7   & $\mathrm{2s^{2}\,2p^{2}(^{3}P)\,3s~^{4}P_{3/2}}$          & 0.82 + 0.12~$\mathrm{2s\,2p^{4}~^{4}P}$                                                                                           & 83366       &    83317.830    &  -48    &  2.42E-09 &  2.42E-09\\                  
8   & $\mathrm{2s^{2}\,2p^{2}(^{3}P)\,3s~^{4}P_{5/2}}$          & 0.81 + 0.12~$\mathrm{2s\,2p^{4}~^{4}P}$                                                                                           & 83413       &    83364.620    &  -49    &  2.40E-09 &  2.39E-09\\                  
9   & $\mathrm{2s^{2}\,2p^{2}(^{3}P)\,3s~^{2}P_{1/2}}$          & 0.95                                                                                                                                               & 86187       &    86137.350    &  -50    &  2.10E-09 &  2.10E-09\\       
10  & $\mathrm{2s^{2}\,2p^{2}(^{3}P)\,3s~^{2}P_{3/2}}$          & 0.95                                                                                                                                               & 86270       &    86220.510    &  -50    &  2.10E-09 &  2.10E-09\\       
... &  ...& ...& ... & ... & ... & ... \\
\hline 
\end{tabular}
\tablecomments{The third column gives wave function composition (up to three \textit{LS}-components with a {fractional} contribution {of} $> 0.02$ of the total wave function) in \textit{LS}-coupling. $\Delta E$ donate the differences between the MCDHF/RCI calculated values and the compiled values from NIST-ASD, i.e. $\Delta E$ = $E_{\text{NIST}} - E_{\text{RCI}}$. {Note that the labelling of the two level pairs, \#74-75 and \#80-82, given in the second column are inverted compared to that of the NIST-ASD data.}}
\end{table*}

\begin{figure}
\centering
\includegraphics[scale=0.45]{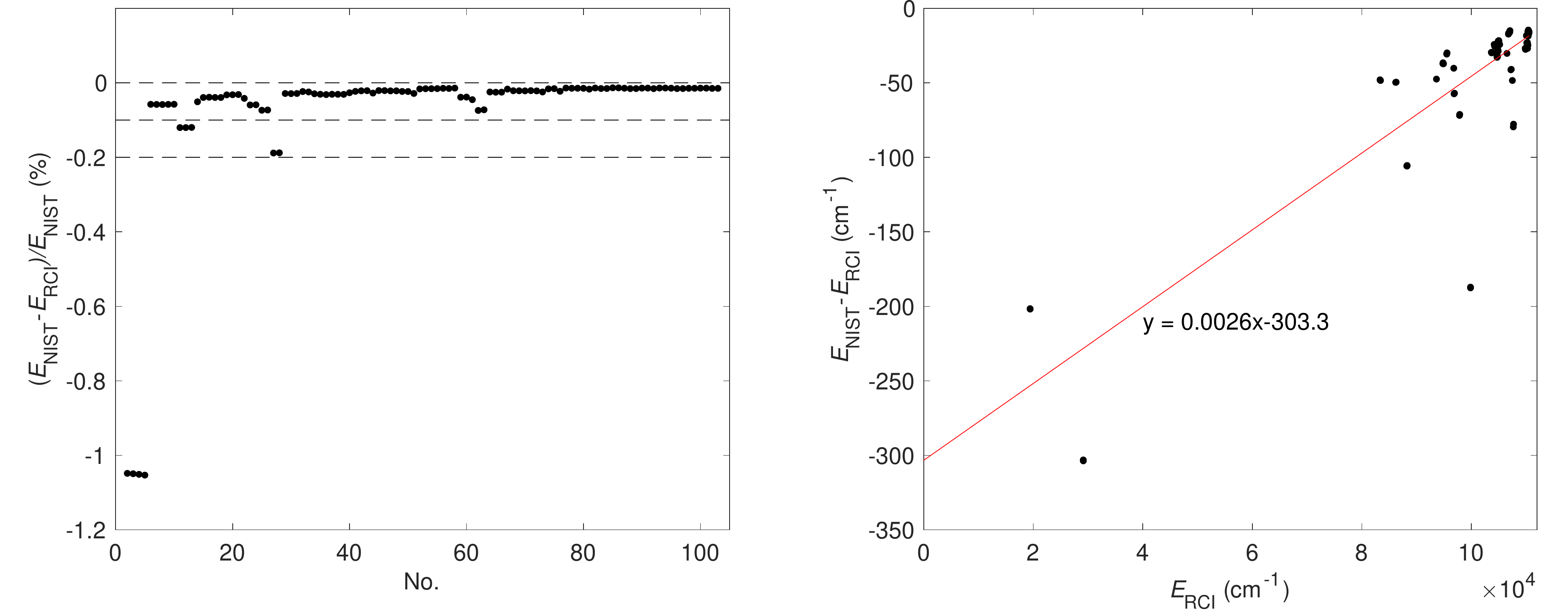}
\caption{Comparison of the current computed excitation energies with the values available in the NIST-ASD. {The left panel shows the relative difference for each level, while the right panel shows the absolute difference along with present computed energies. The dashed lines in the left panel, from upper to lower, indicate the
0.2\%, 0.1\% and 0.0\% relative discrepancies, respectively. The red solid line in the right panel is the linear fit to the scatter data shown in the figure.} }
\label{fig:energy}
\end{figure}

The results of the comparison of computed energy levels in this work with experimental values are shown in Fig.~\ref{fig:energy}. As seen in {the left panel of} Fig.~\ref{fig:energy}, our computed energies agree very well with the NIST compiled experimental energies with an average relative difference of 0.07\%.
The largest relative difference between theory and experiment is 1.05\% for the levels belonging to doublet terms of the $\mathrm{2s^22p^3}$ ground configuration.
%\sout{This is a general computational feature, and such large differences are often found for states of the ground} \sout{configurations belonging to different terms. The large energy differences can be attributed to 
%the relatively slow} \sout{convergence of the captured correlation energy of the doublet terms with respect to the increasing orbital set compared} \sout{to
%the more rapid convergence of the ground quartet term, for which all spins are aligned, leading to an energy} \sout{unbalance, see for example in \cite{mchfbook} (Chapter 5).} %\pj{quote MCHF book chapter 5 and I will find another good reference for this)}.
For the remaining 98 levels, the relative differences of the computed excitation energies are much smaller, with an average of 0.03\%. 
{The right panel of Fig.~\ref{fig:energy} shows the energy differences between NIST-ASD values and the present computed data, $\Delta E = E_{\rm{NIST}} – E_{\rm{RCI}}$, plotted against the excitation energies, $E_\mathrm{RCI}$. We can see that the largest error happens for the ground levels,
which are about 200-300 cm$^{-1}$ higher than the NIST-ASD values.
The root mean square (rms) of the $\Delta E$ values is about 67 cm$^{-1}$.
When the computational excitation energies are corrected with the linear fitting results, the rms of the $\Delta E$ values decreased to 28 cm$^{-1}$, with a systematic error of 0.26\% of the excitation energies. 
%This will make all computed excitation energies smaller. When this computational error is corrected by increasing all excitation energies by 300 cm$^{-1}$, the errors in thus adjusted computed excitation energies become indeed roughly proportional to energies. These adjusted excitation energies of present work have a systematic error of $+0.26\%$ combined with a quasi-random scatter with a rms of about 28 cm $^{-1}$.
}

In Table~\ref{tab:lifetime} we compare the lifetimes from present calculations with other theoretical results and experimental values, when available. 
The calculated lifetimes in Babuskin and Coulomb forms are in agreement within 2.6\%.
Compared with the theoretical results from CIV3 calculations by~\citet{hibbert.1991} and MCHF-BP calculations by~\citet{fischer.2004}, excellent {agreement is} observed for 7 out of 8 lifetime values, with the relative difference between each two of them being less than 7\%. For the $\mathrm{2p^2(^3P)4s~^2P}$ term, the lifetime value from {the} present work is about a factor of two smaller than the CIV3 results, but in agreement with the MCHF-BP value, with the relative difference less than 6\% and in better agreement with the experimental value from~\cite{dumont.1974}.
Good agreement is also observed between computed and experimental lifetime values, except for the 
$\mathrm{2p^2(^3P)3p~^4P^o_{5/2}}$ and $\mathrm{2p^2(^3P)4s~^2P}$ levels.
%Comparing the lifetime values as deduced from this work with various experiments, an excellent agreement is observed for $ 2p^2(^3P)3p~^4S_{3/2}$, $2s2p^4~^4P$, $2p^2(^3P)3s~^4P$, $2p^2(^3P)3s~^2P$ and $2p^2(^1D)3s~^2D$.
%For the $2p^2(^3P)3p~^4D_{7/2}$ level, our length form value is slightly outside both the two experimental values obtained by~\citet{bengtsson.1992} and~\citet{copeland.1987}, respectively. 
For the $\mathrm{ 2p^2(^3P)3p~^4P^o_{5/2}}$ level, all the theoretical lifetimes differ substantially from the experimental values and the two experimental values by \citet{Bromander.1978} and \citet{Desesquelles.1970} differ by 30\% with each other. On the contrary, the theoretical results obtained from three different methods, i.e. MCDHF, MCHF-BP and CIV3, are in good agreement within 8\%. For the $\mathrm{2p^2(^3P)4s~^2P}$ state, all the theoretical values are larger than the experimental result obtained from \cite{dumont.1974}. Our predicted lifetime is in good agreement with that from MCHF-BP calculation \citep{fischer.2004}, while the value from \cite{hibbert.1991} is larger than the others. New experimental measurements of lifetimes would therefore be welcome for these levels.

\begin{table*}[]
\setlength{\tabcolsep}{5.5mm}
\caption{\label{tab:lifetime} Comparison of the experimental and calculated lifetimes (in ns, B = Babuskin form and C = Coulomb form). The lifetime of a term is the mean value of the lifetime corresponding to the different $J$ values. The notation for experimental values, e.g., 44(2) implies $44 \pm 2$.}
\centering
\begin{tabular}{lcccc}
\hline\hline
\multirow{2}{*}{State or term} & \multicolumn{2}{c}{This work} & \multirow{2}{*}{Other calculations}&~\multirow{2}{*}{Experiments}\\
\cline{2-3}
 & B & C &  & \\
\hline
$\mathrm{2p^2(^3P)3p~^4D_{7/2}}$~ & 39.8~ & 40.3 & 37.9$^a$; 37.12$^b$ & 44(2)$^c$; 43(3)$^d$ \\
$ \mathrm{2p^2(^3P)3p~^4S_{3/2}}$~ & 25.3~& 25.5 & 24.1$^a$; 23.26$^b$~&~26.0(1.5)$^c$; 23.3(2.3)$^e$\\
$\mathrm{2p^2(^3P)3p~^4P_{5/2}}$~& 33.4~ & 33.9 & 31.6$^a$; 31.22$^b$& 39$^f$; 55(5)$^g$\\
$\mathrm{2s2p^4~^4P}$ & 6.90 & 6.75 & 6.77$^a$; 7.14$^b$ & 7.3(0.7)$^h$; 7.4(0.4)$^i$; 7.0(0.2)$^j$\\
& & & & 7.1(0.4)$^f$; 5.5(1.5)$^k$, 9.9(1.0)$^l$; 7.2(0.7)$^m$\\
$\mathrm{2p^2(^3P)3s~^4P}$~& 2.42 & 2.42 & 2.48$^a$; 2.59$^b$ & 2.35(0.23)$^{f,h}$; 2.4(0.1)$^j$; 2.2(0.4)$^l$; 2.5(0.3)$^m$\\
$\mathrm{2p^2(^3P)3s~^2P}$~& 2.10 & 2.10 & 2.13$^a$; 1.95$^b$ & 1.9(0.3)$^{h,m}$; 2.3(0.4)$^h$; 2.2(0.1)$^j$\\
& & & & 1.7(0.4)$^l$; 1.9(0.4)$^l$; 2.28(0.2)$^f$\\
$\mathrm{2p^2(^1D)3s~^2D}$~ & 2.57 & 2.58 & 2.44$^a$; 2.44 $^b$ & 2.27(0.3)$^h$; 2.26(0.3)$^h$; 2.65(0.3)$^f$\\
& & & & 2.6(0.1)$^j$; 2.5(0.4)$^k$; 2.2(0.3)$^m$\\
$\mathrm{2p^2(^3P)4s~^2P}$~ & 8.52 & 8.75 & 8.02$^a$; 18.8$^b$ & 6.2$^h$ \\
\hline
\end{tabular}
\tablecomments{$^a$ \citet{fischer.2004}; $^b$~\citet{hibbert.1991}; $^c$~\citet{bengtsson.1992};~$^d$~\citet{copeland.1987}; $^e$~\citet{catherinot.1979};~$^f$~\citet{Desesquelles.1970};~$^g$~\citet{Bromander.1978}; $^h$~\citet{dumont.1974};
~$^i$~\citet{smith.1970.v2};~$^j$~\citet{berry.1971.v61};~$^k$~\citet{mallow.1972.v12};~$^l$~\citet{hutchison.1971.v11};~$^m$~\citet{lawrence.1966.v141}}
\end{table*}

\subsection{Transition parameters}
The E1 transition data, such as wavelengths ($\lambda$), transition rates ($A$), line strengths ($S$), weighted oscillator strengths ($gf$) in Babushkin gauge, are listed in Table~\ref{tab:transition}. {We note that the wavelengths in Table~\ref{tab:transition} are computed from the experimental energy levels compiled in the NIST-ASD and the transition parameters $A$ and $gf$ are also adjusted to the NIST-ASD wavelength values.} 

\begin{table*}[]
\setlength{\tabcolsep}{1.2mm}
\caption{\label{tab:transition}Electric dipole transition data for N I from present calculations. Upper and lower states, wavelength {in vacuum}, $\lambda{_{vac.}}$, transition probability, $A$, line strength, $S$ (in unit au of $a^2_0e^2$), weighted oscillator strength, $\lggf$, together with the relative difference between two gauges of $A$ values, d$T$, and Accuracy class, Acc., {are shown in the table. The wavelengths and all the transition parameters are adjusted to the NIST-ASD Ritz wavelength values. The accuracy class, Acc., is given by: A (Uncertainty $\le$ 3\%), B+ (3\% $<$ Uncertainty $\le$ 7\%), B (7\% $<$ Uncertainty $\le$ 10\%), C+ (10\% $<$ Uncertainty $\le$ 18\%), C (18\% $<$ Uncertainty $\le$ 25\%), D+ (25\% $<$ Uncertainty $\le$ 40\%), D (40\% $<$ Uncertainty $\le$ 50\%), and E (50\% $>$ Uncertainty). This table is available in its entirety in electronic form.}}
\centering
\scriptsize{
\begin{tabular}{cccccccccccc}
\hline\hline
\multirow{3}{*}{Upper} & \multirow{3}{*}{Lower} & \multirow{3}{*}{$\lambda$ (\AA)} &\multirow{3}{*}{$A_B$ (s$^{-1}$)} & \multirow{3}{*}{$A_C$ (s$^{-1}$)} & \multirow{3}{*}{$\lggf_B$} & \multirow{3}{*}{$\lggf_C$} &\multirow{3}{*}{$S_B$ } &\multirow{3}{*}{$S_C$ } &  \multirow{3}{*}{d$T$} & \multicolumn{2}{c}{Acc.} \\
\cline{11-12}
&&&&&&&&&&d$T\&$ & $gf_{\rm{RCI}}\&$\\
&&&&&&&&&&$A$ &$gf_{\rm{NIST-ASD}}$\\
\hline
$\mathrm{2s^2 \,2p^2 (^3P)\,4d~^2D_{ 5/2  }}$      &   $\mathrm{2s^2 \,2p^3~^4S_{ 3/2  }^o}$              &             905.221 &  1.444E+06   &   1.623E+06   &  -2.973   & -2.922   &  3.172E-03   &  3.565E-03     & 0.110   &    C+  &  C   \\
$\mathrm{2s^2 \,2p^2 (^3P)\,4d~^2D_{ 3/2  }}$      &   $\mathrm{2s^2 \,2p^3~^4S_{ 3/2  }^o}$              &             905.411 &  4.649E+05   &   5.202E+05   &  -3.641   & -3.592   &  6.813E-04   &  7.623E-04     & 0.106   &    C+  &  D+  \\
$\mathrm{2s^2 \,2p^2 (^3P)\,4d~^4D_{ 5/2  }}$      &   $\mathrm{2s^2 \,2p^3~^4S_{ 3/2  }^o}$              &             905.786 &  1.657E+07   &   1.861E+07   &  -1.913   & -1.862   &  3.647E-02   &  4.096E-02     & 0.110   &    C+  &  B   \\
$\mathrm{2s^2 \,2p^2 (^3P)\,4d~^4D_{ 3/2  }}$      &   $\mathrm{2s^2 \,2p^3~^4S_{ 3/2  }^o}$              &             905.834 &  3.476E+07   &   3.898E+07   &  -1.767   & -1.717   &  5.101E-02   &  5.719E-02     & 0.108   &    C+  &  B   \\
$\mathrm{2s^2 \,2p^2 (^3P)\,4d~^4P_{ 1/2  }}$      &   $\mathrm{2s^2 \,2p^3~^4S_{ 3/2  }^o}$              &             905.914 &  5.361E+07   &   6.005E+07   &  -1.880   & -1.830   &  3.935E-02   &  4.407E-02     & 0.107   &    C+  &  B   \\
$\mathrm{2s^2 \,2p^2 (^3P)\,4d~^4D_{ 1/2  }}$      &   $\mathrm{2s^2 \,2p^3~^4S_{ 3/2  }^o}$              &             906.207 &  3.201E+07   &   3.588E+07   &  -2.103   & -2.054   &  2.352E-02   &  2.636E-02     & 0.108   &    C+  &  C+  \\
$\mathrm{2s^2 \,2p^2 (^3P)\,4d~^4P_{ 3/2  }}$      &   $\mathrm{2s^2 \,2p^3~^4S_{ 3/2  }^o}$              &             906.432 &  4.903E+07   &   5.510E+07   &  -1.617   & -1.566   &  7.209E-02   &  8.101E-02     & 0.110   &    C+  &  B   \\
$\mathrm{2s^2 \,2p^2 (^3P)\,4d~^2F_{ 5/2  }}$      &   $\mathrm{2s^2 \,2p^3~^4S_{ 3/2  }^o}$              &             906.619 &  2.562E+07   &   2.893E+07   &  -1.722   & -1.670   &  5.655E-02   &  6.385E-02     & 0.114   &    C+  &  B   \\
$\mathrm{2s^2 \,2p^2 (^3P)\,4d~^4P_{ 5/2  }}$      &   $\mathrm{2s^2 \,2p^3~^4S_{ 3/2  }^o}$              &             906.731 &  3.759E+07   &   4.248E+07   &  -1.556   & -1.503   &  8.298E-02   &  9.377E-02     & 0.115   &    C+  &  B   \\
$\mathrm{2s^2 \,2p^2 (^3P)\,4d~^2P_{ 1/2  }}$      &   $\mathrm{2s^2 \,2p^3~^4S_{ 3/2  }^o}$              &             907.069 &  1.240E+06   &   1.416E+06   &  -3.514   & -3.457   &  9.139E-04   &  1.043E-03     & 0.124   &    C+  &  D+  \\
...& ...& ...& ...& ...& ...& ...& ...& ... &... &... &... \\
\hline
\end{tabular}}
\tablecomments{$A_B$, $\lggf_B$, and $S_B$ are, respectively, transition rates, weighted oscillator strengths, and line strengths in the Babushkin (B) form.  $A_C$, $\lggf_C$, and $S_C$ are, respectively,
transition rates, weighted oscillator strengths, and line strengths in the Coulomb (C) form.}
\end{table*}

As mentioned in section~\ref{theory}, the d$T$ values can be used to assess the accuracy of the transition parameters. 
The former are shown in the last column of Table~\ref{tab:transition}. We also investigated the distribution of d$T$ values with respect to the magnitude of the transition rates $A$. 
In Table~\ref{tab:dT}, the transitions are organized in five groups based on the magnitude of the $A$ values and 
the statistical results of d$T$ values, i.e. the mean value $\langle \rm{d}\it{T} \rangle$ and the corresponding standard deviations $\sigma$, are given for each group. {$\langle \rm{d}\it{T} \rangle$ and $\sigma$ are defined as}
\begin{equation}
 \langle \rm{dT} \rangle=\frac{\sum_{\it{i=1}}^{\it{n}}\rm{d}\it{T}_i}{\it{n}}   
\end{equation}
\begin{equation}
\sigma=\sqrt{\frac{\sum_{\it{i=1}}^{\it{n}}(\rm{d}\it{T}_i-\langle \rm{dT} \rangle)^2}{(n-1)}}  
\end{equation}
{where $n$ is the number of transitions of each group.}

As can be seen from Table~\ref{tab:dT}, stronger transitions are usually associated with smaller d$T$ values, while weaker transitions are associated with relatively larger d$T$ values. This is expected since these weaker transitions are mainly unexpected, $LS$-forbidden E1 transitions, for example, the intercombination transitions and two-electron one-photon transitions, and the challenging nature involved in the computations of these types of transitions 
due to extensive cancellation between large contributions, see e.g. \citet{ynnerman.1995}. {The cancellation effect can be represented by cancellation factor (CF). Computed line strengths are expected to show large uncertainties when CF is smaller than about 0.1 or 0.05~\citep{cowan.1981}. For $LS$-allowed transitions computed in this work, 537 out of 905 transitions are associated with CFs large than 0.05, while only 167 out of 796 transitions for $LS$-forbidden transitions.}
For most of the strong transitions with $A > 10^6~ \mbox{s}^{-1}$, the mean $\langle \rm{d}\it{T} \rangle$ is less than {0.05} ($\sigma=0.08$). 
The proportions of the transitions with different d$T$ values are also statistically analysed and shown in the last three rows: they are 84.3\%, 66.0\%, and 51.7\% for transitions with d$T$ less than {0.2, 0.1, and 0.05}, respectively.

\begin{figure}[h!]
\centering
\includegraphics[width=0.63\textwidth,clip]{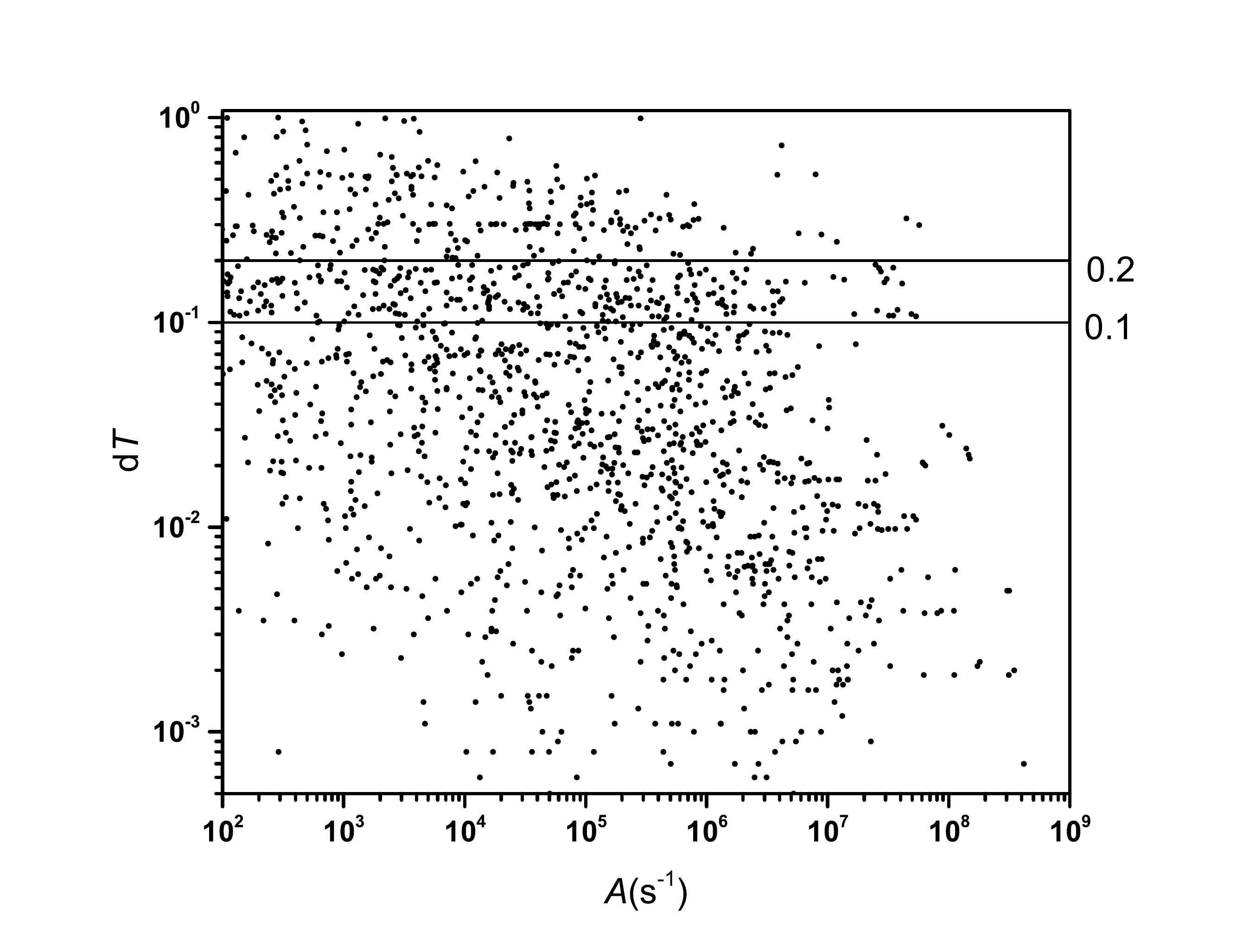}
\caption{Distributions of d$T$ values along with transition rates $A$ in s$^{-1}$. The solid lines indicate the {0.1 and 0.2} relative difference between the Babuskin and Coulomb gauges. {Note that logarithmic scale is used in both x- and y-axis.}}
\label{fig:dT}
\end{figure}

Fig.~\ref{fig:dT} depicts the distributions of the d$T$ values for all the E1 transitions with $A > 10^2~\mbox{s}^{-1}$. Being consistent with the finding in Table~\ref{tab:dT}, that stronger transitions are associated with smaller d$T$ values and the d$T$ values are within {0.2} for most of the transitions. The mean d$T$ for all presented E1 transitions shown in Fig.~\ref{fig:dT} is 0.107 ($\sigma = 0.18$).

\begin{table}[]
\setlength{\tabcolsep}{15mm}
\caption{\label{tab:dT} Statistical results of d$T$ {for} the computed transition rates. The transitions are divided into five groups based on the magnitude of the $A$ values (in s$^{-1}$). The number of transitions (No.), the mean d$T$
($\langle \rm{d}T \rangle$), and the standard deviations ($\sigma$), are given for each group of the transitions. The last three rows show the proportions (in \%) of the transitions with d$T$ less than {0.2, 0.1, and 0.05} in all the transitions with $A \geq 10^2 \mbox{s}^{-1}$. }
\centering
\begin{tabular}{cccc}
\hline\hline
Group & No. &  $\langle \rm{d}\it{T} \rangle$ & $\sigma$  \\
\hline
$< 10^0$ & 68 & 65.77 & 0.35 \\
$ 10^0 - 10^2$ & 199 & 26.65 & 0.25 \\
$10^2 - 10^4$ & 422 & 17.70 & 0.20 \\
$10^4 - 10^6$ & 670 & 9.42 & 0.12 \\
$> 10^6$ & 342 & 4.51 & 0.08 \\
\hline
d$T <20$ & \multicolumn{3}{c}{84.31}\\
d$T < 10$ & \multicolumn{3}{c}{66.04}\\
d$T < 5$ & \multicolumn{3}{c}{51.74}\\
\hline
\end{tabular}
\end{table}

\begin{figure}[h!]
\centering
\includegraphics[width=0.63\textwidth,clip]{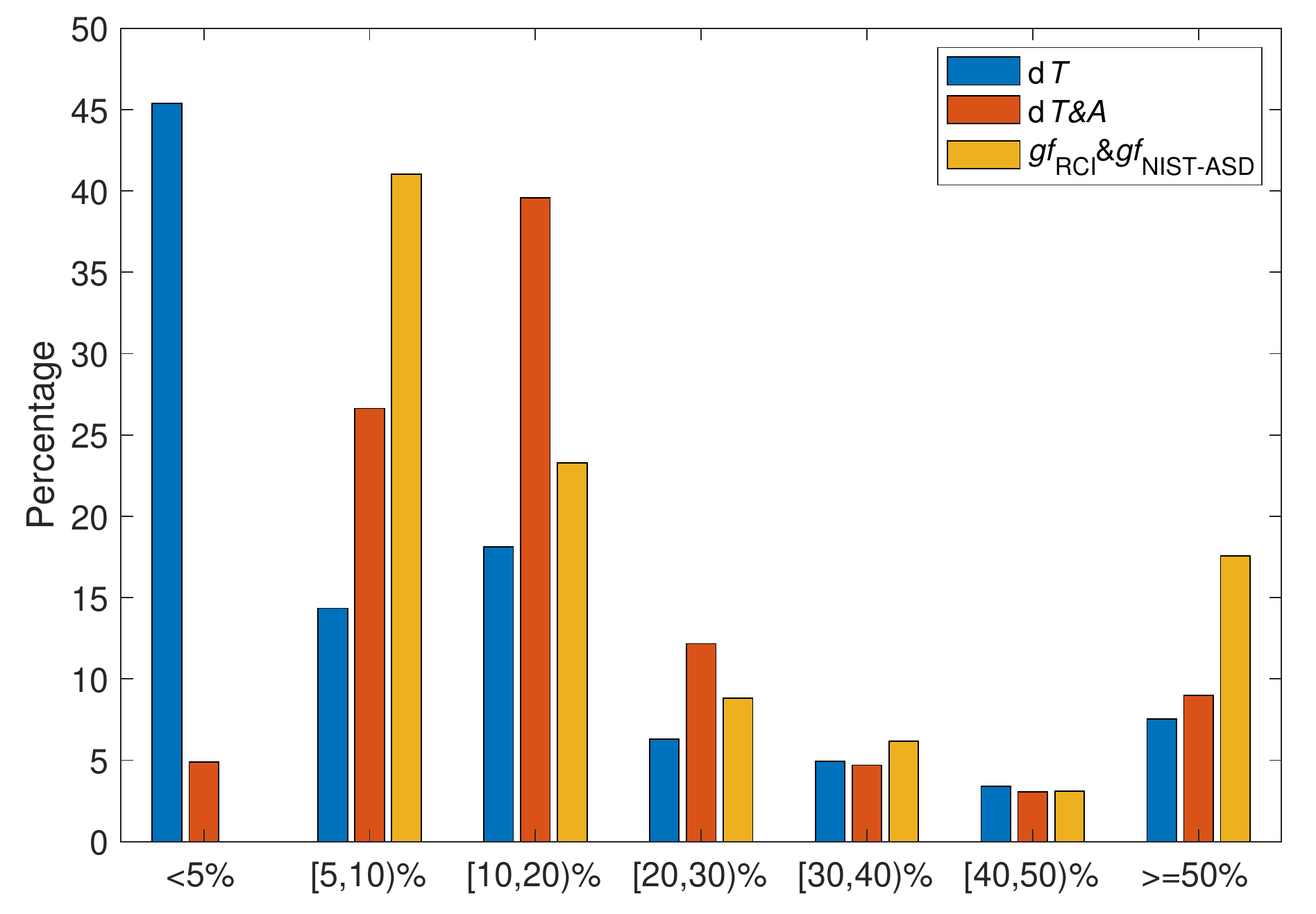}
\caption{Percentage fractions of all transitions in different uncertainty categories, for the uncertainties based on d$T$ values only (blue), d${T}$\&$A$ procedure (red), {and $gf_\mathrm{RCI}$\&$gf_\mathrm{NIST-ASD}$ method (yellow) of \cite{Kramida2013}.}}
\label{fig:accuracy}
\end{figure}

However, the estimation of uncertainties for each transitions is not trivial and there are a number of methods proposed for estimation of uncertainties of 
calculated transition rates \citep{2014Atoms...2...86K,fischer.2009, ekman.2014,ELSAYED2021107930,Gaigalas_2020}. 
In this work, the estimation of uncertainty is performed {by two methods. The first method, which we call d${T}$\&$A$ procedure, is performed in}
steps as follows: 1. The transitions are divided in groups based on the $A$ values, i.e.
$A$ $<$ 10$^{0}$ s$^{-1}$, 
10$^{0}$ $\le A$ $<$ 10$^{2}$ s$^{-1}$, 10$^{2}$ $\le A$ $<$ 10$^{3}$ s$^{-1}$, 10$^{3}$ $\le A$ $<$ 10$^{4}$ s$^{-1}$, 
10$^{4}$ $\le A$ $<$ 10$^{5}$ s$^{-1}$, 10$^{5}$ $\le A$ $<$ 10$^{7}$ s$^{-1}$, and $A \ge$ 10$^{7}$ s$^{-1}$.
2. The averaged d$T_\mathrm{av}$ is determined for each group. 3. The related uncertainty percentage for each transition equals max(d$T$,d$T_\mathrm{av}$). 
{The second method employs the procedure from \cite{Kramida2013}, which evaluates the accuracy of the computed transition probabilities from comparison of $gf_\mathrm{RCI}$ and results from NIST-ASD, $gf_\mathrm{NIST-ASD}$, and we label it as $gf_\mathrm{RCI}$\&$gf_\mathrm{NIST-ASD}$ method.}
The statistical analysis of the number of transitions belonging to specific accuracy class is performed and the percentage fractions {obtained from the above two methods} are shown in Figure \ref{fig:accuracy}.

For comparison, the percentage fractions in different uncertainty categories obtained from d$T$ values are also shown in Figure \ref{fig:accuracy}. 
We can see that the percentage fraction belonging to high-accuracy class is decreased from 45\% by using the d$T$ indicator to 5\% using the d${T}$\&$A$ procedure {and 0 using the $gf_\mathrm{RCI}$\&$gf_\mathrm{NIST-ASD}$ method}, which indicates that using the d$T$ values only for accuracy estimations might underestimate the uncertainties. 
{The accuracy classes obtained from d${T}$\&$A$ and $gf_\mathrm{RCI}$\&$gf_\mathrm{NIST-ASD}$ methods are given for each transition in the last two columns of Table \ref{tab:transition}.}

\begin{figure}
\centering
\includegraphics[width=0.63\textwidth,clip]{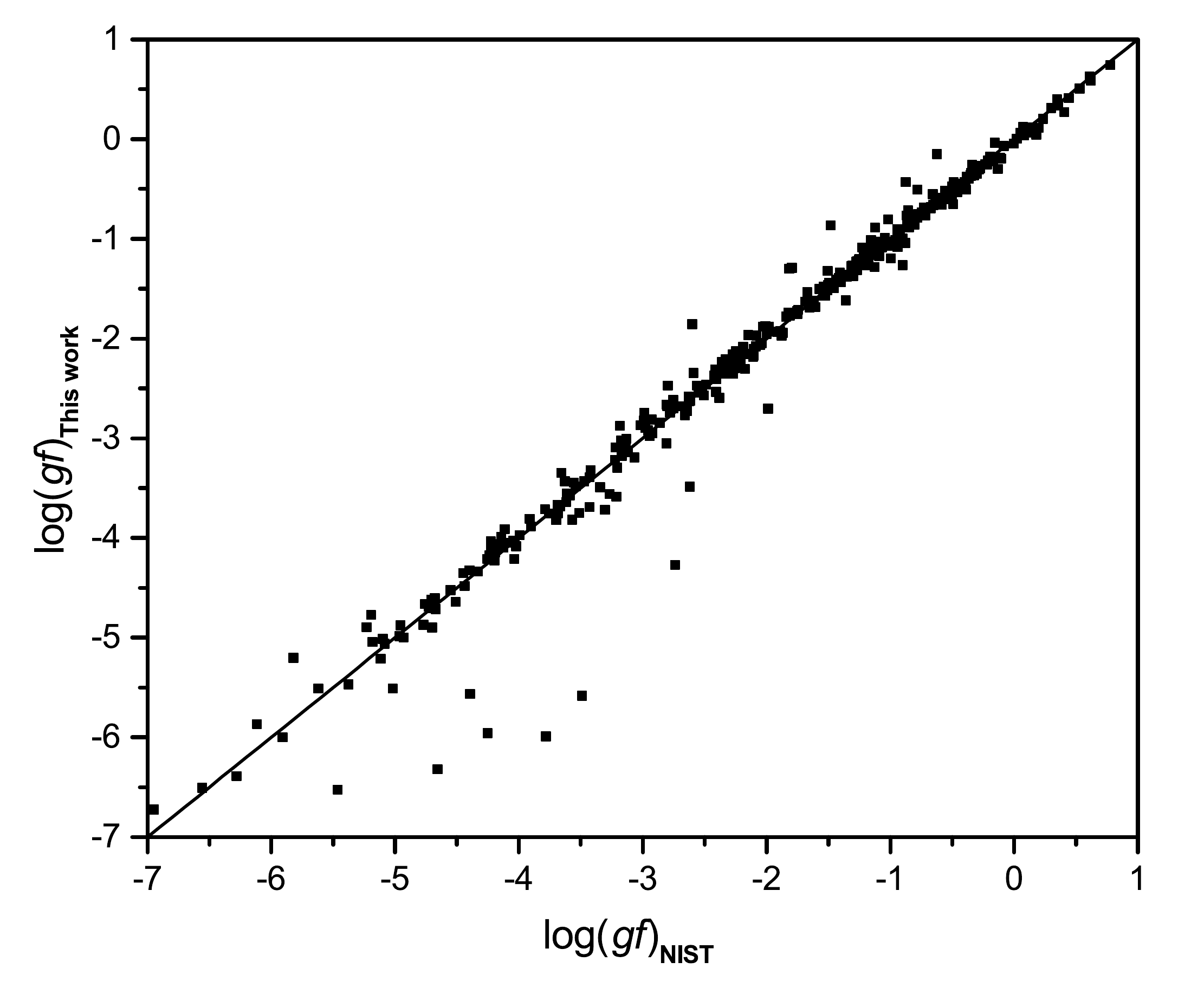}
\caption{Comparison of the log($gf$) values from the present calculations with the results available in the NIST-ASD. }
\label{fig:com_nist}
\end{figure}

The accuracy of computed transition parameters can also be estimated by comparisons with previous calculations and experiments. Fig.~\ref{fig:com_nist} shows the comparison of the present log($gf$) values and the results from the NIST-ASD. 
Note that only the values in the NIST-ASD with uncertainties marked C and above are used for comparison. 
Those NIST-ASD values are compiled by~\citet{wiese.1996} and~\citet{wiese.2007} based on the results from~\citet{tachiev.2002,musielok.1995}, and \citet{hibbert.1991}. 
%\ama{[NIST is just a database, and their data are very similar to what is compared in the following figure, Figure 4.  Is there 
%a need to have two separate figures?  I would suggest to just keep Figure 4, noting that these data are what make up (most of) NIST for N1? \jon{Jon: Sounds like a good idea, keep it simple!}]}
We can see that the agreement between our present computed log($gf$) results and the values from the NIST-ASD {is rather good} for most of the transitions. 

\begin{figure}
\centering
\includegraphics[width=0.63\textwidth,clip]{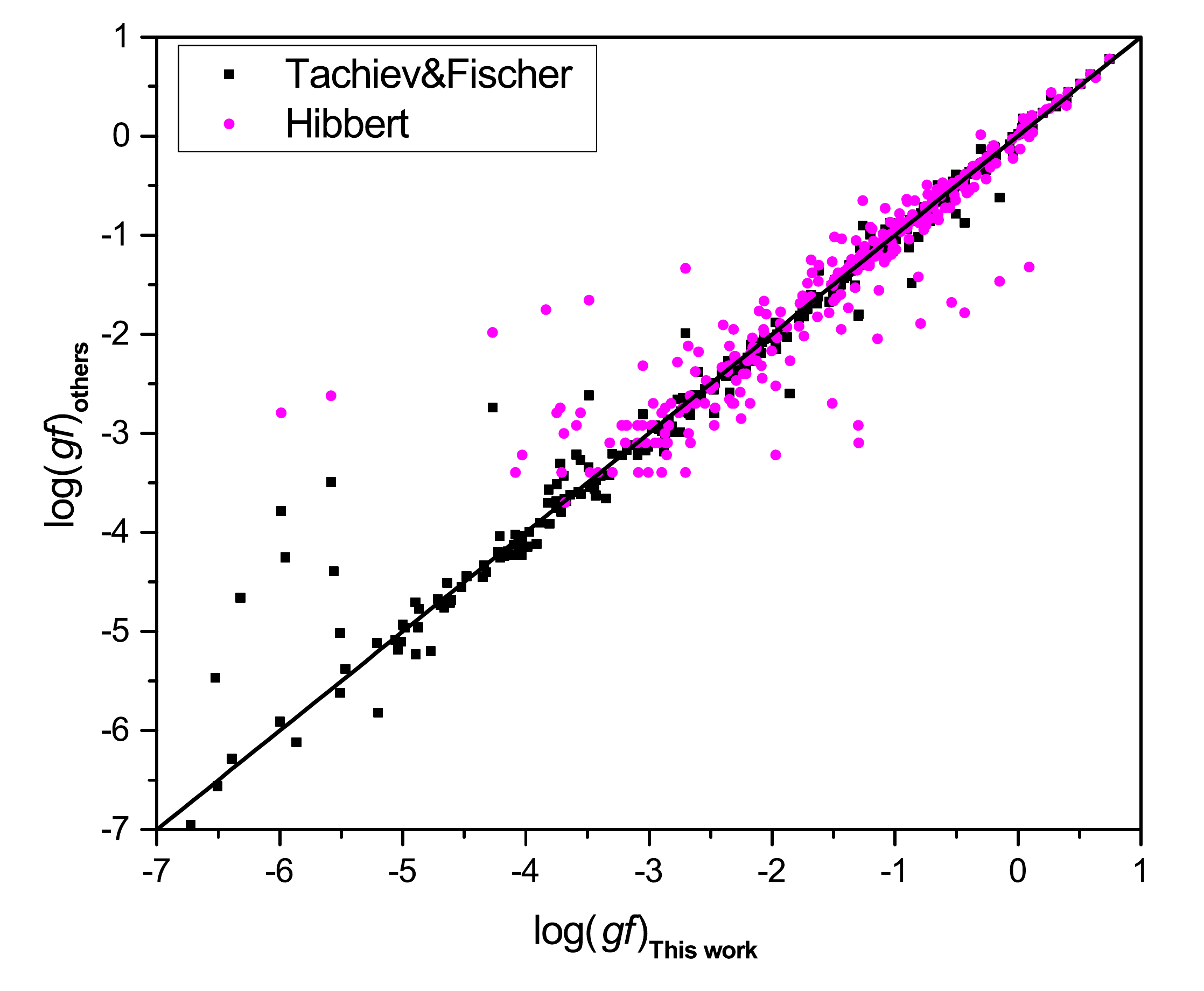}
\caption{Comparison of the log($gf$) values from the present calculations with the values by \citet{tachiev.2002} and ~\citet{hibbert.1991}. }
\label{fig:theory}
\end{figure}

In Fig.~\ref{fig:theory}, we compare our computed log($gf$) values with the results from the other two calculations by \citet{hibbert.1991}, performed with the CIV3 code and \citet{tachiev.2002}, using MCHF-BP method.
As is seen in the figure, the differences between the log($gf$)
values computed in this work and results from other
sources are rather small for most of the transitions.
Comparing the present MCDHF/RCI results with those from MCHF-BP calculations by \citet{tachiev.2002}, which are adopted in the NIST-ASD, 304 (261) out of 334 transitions are in agreement within 25\% (10\%). The exceptions are the weak transitions with log($gf$) $< -5$ . 181 out of 305 transitions are in agreement within 25\% with the results from \citet{hibbert.1991}.
It seems that when both theoretical results, i.e. from \citet{hibbert.1991} and \citet{tachiev.2002}, are available for the transitions, the present MCDHF/RCI results are in better agreement with the latter.

In Table~\ref{tab:com}, the computed transition probabilities are compared with the experimental results from \citet{musielok.1995} and \citet{bridges.2010}.
%The two experimental data are normalized to the experimental lifetime of $2p^2(^3P)3p~^4D_{7/2}$ and $2p^2(^3P)3p~^4S_{3/2}$ levels, measured by~\citet{bengtsson.1992}, \citet{copeland.1987} and~\citet{catherinot.1979}. In case of this, direct comparisons may not be suitable. 
The theoretical results obtained from \citet{hibbert.1991} and \cite{tachiev.2002}, when available, are also listed for comparison. The estimated uncertainties d$T$ of the MCDHF/RCI transition rates are given in parentheses. 
From Table~\ref{tab:com} we can seen that when the two experimental results from \citet{musielok.1995} and \citet{bridges.2010} are both available for the multiplet, the present results are in slightly better agreement with the latter for the $\mathrm{2p^2(^3P)3p\rightarrow2p^2(^3P)3s}$ transitions, while the situation is getting complex in the case of $\mathrm{2p^2(^3P)3d\rightarrow2p^2(^3P)3p}$ transitions. 
For allowed $\mathrm{2p^2(^3P)3p\rightarrow~2p^2(^3P)3s}$ transitions, present results are in excellent agreement with the two experimental results, as well as with the MCHF-BP~\citep{tachiev.2002} and CIV3~\citep{hibbert.1991} data.
While for the $\mathrm{2p^2(^3P)3d~^4P\rightarrow2p^2(^3P)3p~^4D^o}$ transitions, 
the present results are in better agreement with the experimental values from~\citet{bridges.2010} than those from \citet{musielok.1995}. 
{For the transitions from $\mathrm{2p^2(^3P)3d~^4P_{5/2}}$ and $\mathrm{^2F_{5/2}}$ levels, the transition rates from the present calculations differ substantially from the experimental values by~\citet{bridges.2010}, that is, by 70\%, while the values from  \citep{tachiev.2002} appear to be in better agreement with the corresponding  experimental results. A closer inspection of the wave function composition given in Table \ref{tab:energy} {reveals a strong mixing between these two levels}. It is very difficult to
accurately calculate the radiative properties of such strongly interacting levels.}
\startlongtable
\begin{deluxetable*}{ccccccrc}
\tablecaption{\label{tab:com} Comparison of transition rates obtained from experimental values with present MCDHF/RCI and other calculations. The present values from MCDHF/RCI calculations are given in the Babushkin gauge. The values in the
parentheses are the relative differences between the Babushkin and Coulomb gauges.  
Note that the values from MCHF calculations are those compiled in the NIST database.  {The numbers in parentheses without \% are the uncertainties with respect to the last digit
quoted, for example, 25.31(114) implies $25.31\pm1.14$.}
} 
\setlength{\tabcolsep}{2.5mm}
%\tabletypesize{\small}
\tablehead{
\colhead{} &  \colhead{} &  \colhead{}  & \multicolumn{5}{c}{Transition rate $A$ ($10^6$ s$^{-1}$)}  \\ 
\cline{4-8}
\colhead{Transition array}& \colhead{Term}& \colhead{$g_u - g_l$}& \colhead{MCDHF/RCI} & \colhead{Tachiev}   &\colhead{Hibbert}   &  \colhead{Musielok}  &  \colhead{Bridges}  \\
\colhead{}& \colhead{}& \colhead{}&\colhead{} & \colhead{\& {Fischer}$^a$} & \colhead{et al.$^b$} & \colhead{et al.$^c$} & \colhead{\& Wiese$^d$} 
}
\startdata
$\mathrm{2p^2(^3P)3p-2p^2(^3P)3s}$   &   $\mathrm{^4D^o-^4P}$	 &     8 - 6	  	&  23.95(1.30\%)       &   25.31(114)    &    25.96   &    $ 22.1(24)   $    &    23.2(19)    \\
		                                 &                         &     6 - 4	  	&  17.85(1.30\%)       &   18.81(88)     &    19.25   &    $ 16.7(18)   $    &    17.29(138)   \\
		                                 &                         &     4 - 2	  	&  10.95(1.29\%)       &   11.52(55)     &    11.79   &    $ 10.4(11)   $    &    10.67(85)   \\
		                                 &                         &     6 - 6	  	&  6.167(1.25\%)       &   6.546(351)    &    6.725   &    $ 5.95(71)   $    &    5.96(48)    \\
		                                 &                         &     4 - 4	  	&  12.19(1.27\%)       &   12.88(62)     &    13.20   &    $ 11.7(13)   $    &    12.04(96)   \\
		                                 &                         &     2 - 2	  	&  20.51(1.27\%)       &   21.61(97)     &    22.11   &    $ 19.7(22)   $    &    20.0(16)    \\
		                                 &                         &     4 - 6	  	&  0.907(1.21\%)       &   0.965(59)     &    0.993   &    $ 0.88(12)   $    &                \\
		                                 &                         &     2 - 4	  	&  3.554(1.24\%)       &   3.761(194)    &    3.857   &    $ 3.24(39)   $    &    3.70(30)    \\
$\mathrm{2p^2(^3P)3p-2p^2(^3P)3s}$   &   $\mathrm{^4P^o-^4P}$	 &     6 - 6	  	&  21.43(1.69\%)       &   22.63(155)    &    23.07   &    $ 22.1(27)   $    &                \\
		                        &                                  &     4 - 4	  	&  4.975(1.64\%)       &   5.233(365)    &    5.312   &    $ 5.02(60)    $    &                \\
		                        &                                  &     2 - 2	  	&  4.424(1.73\%)       &   4.684(314)    &    4.771   &    $ 4.85(58)    $    &                \\
		                        &                                  &     4 - 6	  	&  12.40(1.71\%)       &   13.12(92)     &    13.41   &    $ 13.4(16)    $    &                \\
		                        &                                  &     2 - 4	  	&  24.79(1.69\%)       &   26.19(175)    &    26.67   &    $ 26.5(32)    $    &                \\
		                        &                                  &     6 - 4	  	&  7.751(1.74\%)       &   8.220(566)    &    8.385   &    $ 7.86(94)    $    &                \\
		                        &                                  &     4 - 2	  	&  11.84(1.71\%)       &   12.53(84)     &    12.74   &    $ 12.2(15)    $    &                \\
$\mathrm{2p^2(^3P)3p-2p^2(^3P)3s}$   &    $\mathrm{^4S^o-^4P}$ &     4 - 6	  	&  18.34(0.99\%)       &   19.61(94)     &    20.17   &    $ 19.0(21)    $    &    17.91(143)  \\
		                        &                                  &     4 - 4	  	&  11.20(0.96\%)       &   11.96(52)     &    12.36   &    $ 12.3(14)    $    &    11.83(85)   \\
		                        &                                  &     4 - 2	  	&  5.295(0.94\%)       &    5.649(229)   &    5.856   &    $ 5.63(62)    $    &    5.71(46)    \\
$\mathrm{2p^2(^3P)3p-2p^2(^3P)3s}$   &   $\mathrm{^2D^o-^2P}$	 &     6 - 4	  	&  25.73(0.03\%)       &   25.19(170)    &    26.79   &    $ 25.5(31)    $    &                \\
		                        &                                  &     4 - 2	  	&  21.87(0.03\%)       &   21.39(154)    &    22.63   &    $ 21.9(26)    $    &                \\
		                        &                                  &     4 - 4	  	&  3.820(0.04\%)       &   3.741(250)    &    4.043   &    $ 3.88(50)    $    &                \\
$\mathrm{2p^2(^3P)3p-2p^2(^3P)3s}$   &   $\mathrm{^2P^o-^2P}$	 &     4 - 4	  	&  26.38(0.35\%)       &   26.78(32)     &    27.52   &    $ 25.2(30)   $    &                \\
		                        &                                  &     2 - 2	  	&  20.63(0.37\%)       &   21.45(00)     &    21.58   &    $ 19.8(24)   $    &                \\
		                        &                                  &     2 - 4	  	&  10.58(0.32\%)       &   10.75(12)     &    11.08   &    $ 9.70(116)  $    &                \\
		                        &                                  &     4 - 2	  	&  4.772(0.37\%)       &   4.871(42)     &    5.106   &    $ 4.64(60)   $    &                \\
$\mathrm{2p^2(^3P)4p-2p^2(^3P)3s}$   &   $\mathrm{^2P^o-^2P}$  &     4 - 4	  	&  1.024(9.36\%)       &                 &            &    $ 0.749(105) $    &                \\
	                          &                                  &     2 - 2	  	&  0.895(9.61\%)       &                 &            &    $ 0.672(94)  $    &                \\
	                          &                                  &     2 - 4	  	&  0.368(8.63\%)       &                 &            &    $ 0.269(40)  $    &                \\
	                          &                                  &     4 - 2	  	&  0.207(9.67\%)       &                 &            &    $ 0.240(36)  $    &                \\
$\mathrm{2p^2(^3P)3d-2p^2(^3P)3p}$   &   $\mathrm{^2P-^2S^o}$	 &     4 - 2	  	&  26.19(1.19\%)       &   32.08(590)    &    32.72   &    $ 28.7(37)   $    &                \\
		                        &                                  &     2 - 2	  	&  26.02(1.27\%)       &   31.97(583)    &    32.65   &    $ 30.0(39)   $    &                \\
$\mathrm{2p^2(^3P)4d-2p^2(^3P)3p}$   &   $\mathrm{^2P-^2S^o}$	 &     4 - 2	  	&  4.054(12.61\%)      &                 &            &    $ 3.58(50)   $    &                \\
		                        &                                  &     2 - 2	  	&  4.221(12.95\%)      &                 &            &    $ 3.64(51)   $    &                \\
$\mathrm{2p^2(^3P)3d-2p^2(^3P)3p}$   &   $\mathrm{^4F-^4D^o}$  &    10 - 8  	  &  36.11(0.98\%)       &   38.96(315)    &    39.05   &    $ 37.5(49)   $    &                \\
		                        &                                  &     8 - 6	  	&  31.62(0.98\%)       &   34.11(280)    &    34.17   &    $ 31.9(42)   $    &                \\
		                        &                                  &     6 - 4	  	&  27.94(0.97\%)       &   30.15(262)    &    30.17   &    $ 28.5(37)   $    &    36.9(55)    \\
		                        &                                  &     4 - 2	  	&  25.92(0.98\%)       &   28.05(237)    &    27.94   &    $ 26.2(34)   $    &                \\
		                        &                                  &     8 - 8	  	&  3.479(0.90\%)       &   3.833(427)    &    3.991   &    $ 3.97(52)   $    &    4.42(44)    \\
		                        &                                  &     6 - 6	  	&  6.663(0.99\%)       &   7.317(782)    &    7.606   &    $ 7.31(95)   $    &    7.58(68)    \\
		                        &                                  &     4 - 4	  	&  9.067(0.96\%)       &   9.833(782)    &    9.936   &    $ 9.89(129) $    &    10.3(9)     \\
		                        &                                  &     6 - 8	  	&  0.218(0.01\%)       &    0.239(29)     &    0.237   &    $ 0.326(46)  $    &    0.256(31)   \\
		                        &                                  &     4 - 6	  	&  0.518(0.80\%)       &    0.564(51)    &    0.579   &    $ 0.413(58)  $    &                \\
$\mathrm{2p^2(^3P)3d-2p^2(^3P)3p}$   &   $\mathrm{^4P-^4D^o}$  &     6 - 8	  	&  0.570(6.85\%)       &   0.822(169)    &    0.668   &    $ 1.39 (18)  $    &    0.842(109)  \\
		                        &                                  &     4 - 6	  	&  0.179(13.75\%)      &   0.223(102)    &    0.048   &    $ 0.450(63)  $    &    0.236(35)   \\
		                        &                                  &     2 - 4	  	&  0.150(13.72\%)      &    0.183(98)    &            &    $ 0.311(44)  $    &    0.152(23)   \\
		                        &                                  &     6 - 6	  	&  0.606(2.53\%)       &   1.394(850)    &    2.471   &    $ 2.28(30)   $    &    1.304(130)  \\
		                        &                                  &     4 - 4	  	&  1.845(2.37\%)       &   2.015(518)    &    2.787   &    $ 3.64(47)   $    &    2.15(19)     \\
		                        &                                  &     2 - 2	  	&  1.649(3.59\%)       &   1.828(484)    &    2.999   &    $ 2.93(38)   $    &    1.91(25)    \\
		                        &                                  &     6 - 4	  	&  0.529(0.25\%)       &   0.348(179)    &    0.332   &    $ 0.760(106) $    &    0.406(45)   \\
		                        &                                  &     4 - 2	  	&  0.358(1.97\%)       &   0.407(117)    &    0.707   &    $ 0.758(106) $    &    0.407(41)   \\
$\mathrm{2p^2(^3P)3d-2p^2(^3P)3p}$   &   $\mathrm{^4D-^4D^o}$	 &     8 - 8	  	&  9.264(1.29\%)       &   10.28(124)    &    10.02   &    $ 8.82(115)  $    &                \\
		                        &                                  &     6 - 6	  	&  5.149(1.75\%)       &   5.730(704)    &    4.900   &    $ 4.79(62)  $    &                \\
		                        &                                  &     4 - 4	  	&  2.963(2.20\%)       &   3.330(526)    &    2.408   &    $ 2.99(39)  $    &                \\
		                        &                                  &     2 - 2	  	&  3.366(2.09\%)       &   3.829(610)    &    2.640   &    $ 3.22(42)  $    &                \\
		                        &                                  &     6 - 8	  	&  2.675(0.07\%)       &   2.965(404)    &    3.243   &    $ 2.58(34)  $    &                \\
		                        &                                  &     4 - 6	  	&  4.030(0.32\%)       &   4.496(540)    &    4.717   &    $ 4.10(53)  $    &                \\
		                        &                                  &     2 - 4	  	&  4.821(0.76\%)       &   5.414(621)    &    5.603   &    $ 4.77(62)  $    &                \\
		                        &                                  &     8 - 6	  	&  0.577(1.30\%)       &   0.699(111)    &    0.751   &    $ 0.534(75) $    &                \\
		                        &                                  &     6 - 4	  	&  0.989(1.66\%)       &   1.179(185)    &    1.115   &    $ 1.09(14)  $    &                \\
		                        &                                  &     4 - 2	  	&  1.194(1.89\%)       &   1.40(23)      &    1.151   &    $ 1.16(15)  $    &                \\
$\mathrm{2p^2(^3P)3d-2p^2(^3P)3p}$    &  $\mathrm{^4P-^4P^o}$	 &     6 - 6	  	&  2.976(1.05\%)       &   3.928(159)    &    3.069   &    $ 2.37(31)  $    &                \\
	                          &              	                   &     2 - 2	  	&  5.353(0.57\%)       &   5.397(924)    &    9.480   &    $ 3.39(44)  $    &                \\
	                          &              	                   &     4 - 6	  	&  3.721(0.79\%)       &   3.757(363)    &    2.740   &    $ 2.29(30)  $    &                \\
	                          &              	                   &     2 - 4	  	&  6.649(0.89\%)       &   6.751(701)    &    4.178   &    $ 4.01(52)  $    &                \\
	                          &              	                   &     6 - 4	  	&  4.829(0.65\%)       &   7.16(160)     &    10.01   &    $ 4.18(54)  $    &                \\
	                          &              	                   &     4 - 2	  	&  8.942(0.70\%)       &   9.027(793)    &    11.03   &    $ 5.55(72)  $    &                \\
$\mathrm{2p^2(^3P)3d-2p^2(^3P)3p}$   &  $\mathrm{^4D-^4P^o}$	 &     8 - 6	  	&  24.10(0.27\%)       &   25.46(66)     &    25.11   &    $ 23.5(31)  $    &                \\
	                          &                                  &     6 - 4	  	&  12.38(0.18\%)       &   13.20(80)     &    11.08   &    $ 12.3(16)  $    &                \\
	                          &                                  &     4 - 2	  	&  6.058(0.10\%)       &   6.541(706)    &    4.249   &    $ 6.07(79)  $    &                \\
	                          &                                  &     6 - 6	  	&  11.88(0.43\%)       &   12.35(66)     &    13.40   &    $ 11.3(15)  $    &                \\
	                          &                                  &     4 - 4	  	&  15.14(0.36\%)       &   15.91(42)     &    15.40   &    $ 15.2(20)  $    &                \\
	                          &                                  &     2 - 2	  	&  17.97(0.25\%)       &   19.11(120)    &    14.53   &    $ 17.8(23)  $    &                \\
	                          &                                  &     4 - 6	  	&  3.254(0.48\%)       &   3.332(347)    &    4.332   &    $ 3.24(42)  $    &                \\
	                          &                                  &     2 - 4	  	&  7.305(0.42\%)       &   7.553(747)    &    10.10   &    $ 7.16(93)  $    &                \\
$\mathrm{2p^2(^3P)5s-2p^2(^3P)3p}$   &   $\mathrm{^4P-^4P^o}$	 &     6 - 6	  	&  1.811(8.10\%)       &                 &            &    $ 1.75(25)  $    &                \\
		                        &                                  &     4 - 4	  	&  0.392(7.87\%)       &                 &            &    $ 0.445(67) $    &                \\
		                        &                                  &     2 - 2	  	&  0.348(7.72\%)       &                 &            &    $ 0.365(55) $    &                \\
		                        &                                  &     4 - 6	  	&  0.917(7.44\%)       &                 &            &    $ 0.937(141)$    &                \\
		                        &                                  &     2 - 4	  	&  1.838(7.46\%)       &                 &            &    $ 1.94(29)  $    &                \\
		                        &                                  &     6 - 4	  	&  0.758(8.37\%)       &                 &            &    $ 0.738(111)$    &                \\
		                        &                                  &     4 - 2	  	&  1.019(8.08\%)       &                 &            &    $ 0.979(147)$    &                \\
$\mathrm{2p^2(^3P)5s-2p^2(^3P)3p}$   &  $\mathrm{^2P-^2D^o}$   &     4 - 6	  	&  2.252(3.43\%)       &                 &            &    $ 1.79(25)  $    &                \\
		                        &                                  &     2 - 4	  	&  3.086(7.70\%)       &                 &            &    $ 2.85(40)  $    &                \\
		                        &                                  &     4 - 4	  	&  0.193(1.44\%)       &                 &            &    $ 0.230(35) $    &                \\
$\mathrm{2p^2(^3P)3d-2p^2(^3P)3p}$   &   $\mathrm{^2F-^4D^o}$  &     6 - 6	  	&  1.719(0.57\%)       &   1.064(481)    &    0.420   &    $ 2.26(32)  $    &    1.34(16)    \\
		                        &                                  &     8 - 6	  	&  0.595(1.00\%)       &   0.609(331)    &    0.527   &    $ 1.08(15)  $    &    0.646(78)   \\
		                        &                                  &     6 - 4	  	&  0.268(2.32\%)       &   0.485(687)    &    0.555   &    $ 0.810(113)$    &    0.462(55)   \\
$\mathrm{2p^2(^3P)3d-2p^2(^3P)3p}$   &   $\mathrm{^2F-^4P^o}$	 &     6 - 4	  	&  3.564(0.62\%)       &   1.289(1182)   &    0.160   &    $ 1.03(14)  $    & \\
\enddata
\tablecomments{$^a$~\cite{tachiev.2002}; $^b$~\cite{hibbert.1991}; $^c$~\cite{musielok.1995}; $^d$~\cite{bridges.2010}.}
\end{deluxetable*}

%Fig.~\ref{fig:com_exp_theo} shows the comparison of the present transition rates with the experimental values~\citep{musielok.1995} and the other two theoretical results \citep{hibbert.1991,tachiev.2002}. As shown in the figure, the agreement are quite good for most of the strong transitions, but with a much wider scatter for the weaker lines. 
As can be seen from Table~\ref{tab:com} %and Fig.~\ref{fig:com_exp_theo}
, when three theoretical values are available for the transitions, the present MCDHF/RCI results seem to be in better agreement with the experimental values obtained by \citet{musielok.1995} than the others. 50 out of 72 transitions from present calculations agree with~\citet{musielok.1995} within 10\%, while 33(37) out of 72 transitions are within the same range for theoretical data from \citet{hibbert.1991}(\citet{tachiev.2002}).

\begin{figure}
\centering
\includegraphics[width=0.63\textwidth,clip]{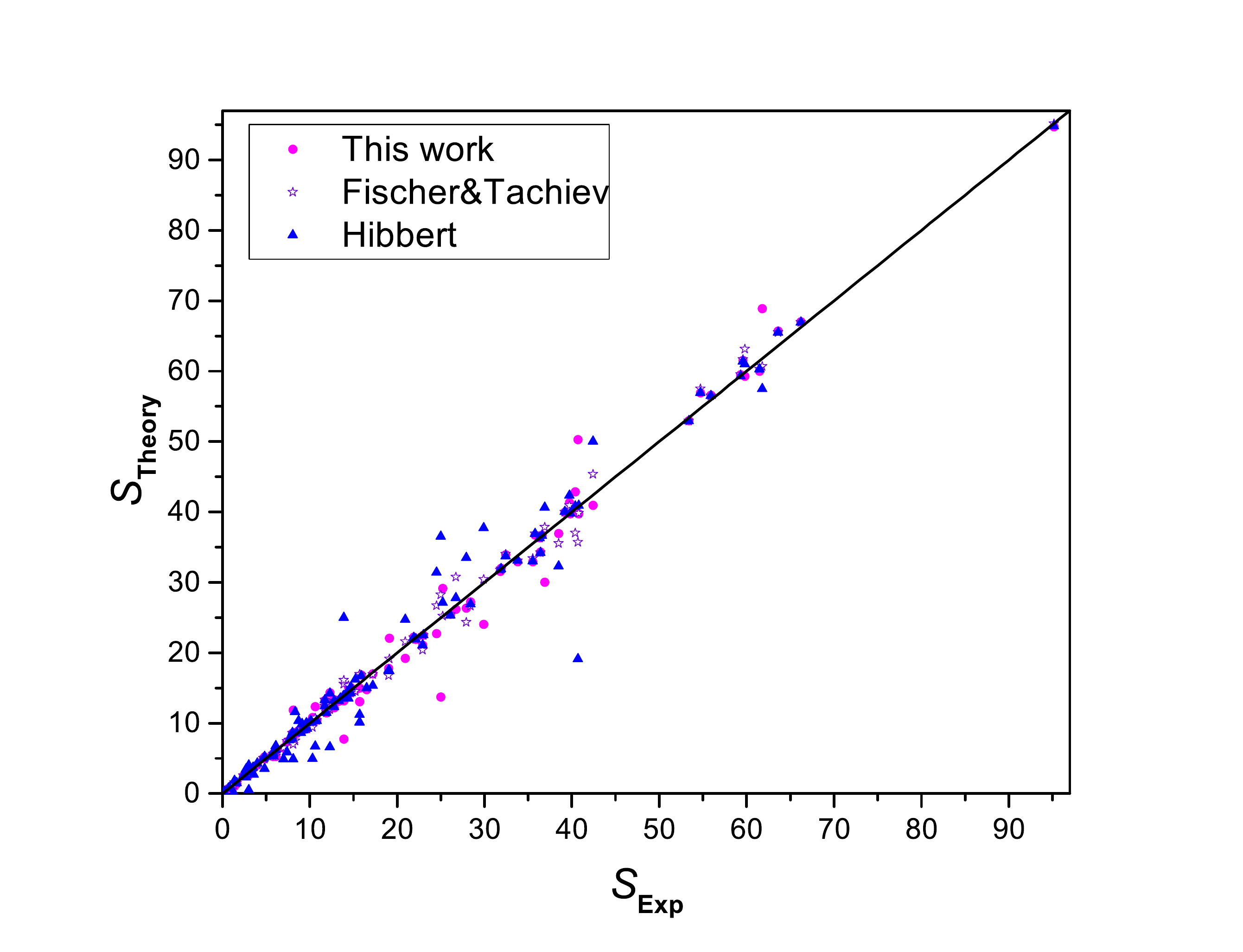}
\caption{Comparison of the theoretical {relative} line strengths, $S$, with the corresponding experimental results published by~\citet{baclawski.2002} and \citet{baclawski.2008}.}
\label{fig:com_s}
\end{figure}

There are also a number of measurements of line strength $S$. In Fig.~\ref{fig:com_s}, the selected computed $S$ values are compared with experimental results by~\citet{baclawski.2002} and~\citet{baclawski.2008}. The theoretical results of \citet{hibbert.1991} and~\citet{fischer.2004} are also shown in the figure. 
{Note that \citet{baclawski.2002} and \citet{baclawski.2008} provided the relative line strengths within multiplets (normalized to 100). Therefore, all the 
%Due to the line strength provided by \citet{baclawski.2002} and~\citet{baclawski.2008} both are the relative line strengths within multiplets (normalized to the sum of 100), so all the
theoretical values used for comparison in Fig.~\ref{fig:com_s} are the fractions of line strength for each transition in a multiplet, {so that} their sum is 100 for each multiplet.}
The computed $S$ values from present work and the MCHF-BP calculations by~\citet{fischer.2004} are in much better agreement with those experimental values than the CIV3 calculations by~\citet{hibbert.1991}. %\pj{than.... are you missing something here?}
The $S$ values from the present work are within the uncertainties of experimental measurements for most of the transitions. 
%The main exceptions are the two transitions which belong to $\mathrm{2p^2(^3P)4s~^2P\rightarrow2p^2(^3P)3p~ ^2D^o}$ multiplet for which the
%present $S$ values differ substantially from the experiment values.
%However, as shown in the figure, the results from the three calculations agree well with each other, therefore the experimental values may need to be re-investigated for this multiplet.

\subsection{Validation via the solar spectrum}

The Sun is our best understood star, 
having well constrained parameters 
\citep{2016AJ....152...41P},
and for which exist both high quality observations of the emergent intensity
\citep{1995ASPC...81...32D,1999SoPh..184..421N,2016A&A...590A.118D},
as well as realistic simulations of the photospheric layer
\citep{2009LRSP....6....2N,2013A&A...554A.118P}.
As such, analyses of the solar intensity spectrum present an alternative way to test
the accuracy of the present calculations.

For this purpose we consider the study of \citet{amarsi.2020}. The authors
presented a detailed analysis of 5 \ion{N}{1} lines
in the solar disc-center intensity spectrum, using state-of-the-art model atmospheres
and radiative transfer methods.
From this analysis they measured, for each line as well as the average over all lines,
the solar nitrogen abundance,
conventionally reported as the logarithmic number density of nitrogen nuclei
relative to that of hydrogen nuclei (plus an offset):
$\lgeps{}\equiv\log_{10}\left(N_{\mathrm{N}}/N_{\mathrm{H}}\right)+12$.
They adopted the transition probabilities given by \citet{tachiev.2002}.
To first order, 
their results can be modified to account for new atomic data using
\begin{equation}
\centering
\lgeps{}_{\text{new}}=\lgeps{}_{\text{Tachiev}}+\Delta_{\text{new}}
\label{eq:abcor}
\end{equation}
where the correction term $\Delta_{\text{new}}$ is given by
\begin{equation}
\centering
\Delta_{\text{new}}=\lggf_{\text{Tachiev}}-\lggf_{\text{new}}\,.
\label{eq:delta}
\end{equation}

Table~\ref{tab:abundance} presents the log($gf$) values of the five permitted lines in \ion{N}{1} used by \citep{amarsi.2020}. 
The log($gf$) values in both Coulomb and Babuskin forms from present calculations are given in the table. We can seen that for these five transition lines, the present log($gf$) values calculated in the Coulomb and Babuskin agree very well, with the d$T$ values below {0.02}.
{We also investigated the convergence of the wavelengths and oscillator strengths between the results from the calculations of the last two layers. The wavelengths {have converged} to within 0.1\% for all the five transitions and the difference between the oscillator strengths from the calculation of the last two layers is 0.0002 for 744.229 nm, 0.01 for 821.633 nm and 868.34 nm, 0.001 for 862.923 nm, and 0.005 for 1010.89 nm.}
Experimental measurements of transition probabilities for the lines were reported by \cite{musielok.1995} and \cite{bridges.2010}; the derived log($gf$) values are shown in the last two columns in Table~\ref{tab:abundance}.
In all cases, our theoretical values and those from \citet{tachiev.2002} fall into the range of the estimated uncertainties of the experimental values.
%The transition parameters from~\citet{tachiev.2002} are compiled in the NIST ASD and are assigned an accuracy level of ``$\rm{B}^+$" ($< 7\%$), except for the $2p^2(^3P)3d~ ^4F_{5/2}\rightarrow2p^2(^3P)3p~ ^4D_{3/2}^o$ transition with an accuracy of ``B" , $< 10\%$.
The previous calculation by \citet{hibbert.1991} yielded larger log($gf$) values than the experimental determinations for 821.633 nm, 868.340 nm lines while the recommended value for 868.340 nm line by \cite{2022arXiv220614095B} is slightly larger than the experimental results.
\begin{table*}
\centering
\caption{\label{tab:abundance} Comparison of the log($gf$) values of the five lines used for abundance analysis in~\citet{amarsi.2020}. B = Babuskin form and C = Coulomb form.}
\scriptsize
\begin{tabular}{cccrrrrrrr}
\hline\hline
 &  &  & \multicolumn{7}{c}{log($gf$)} \\
\cline{4-10}
Upper &  Lower &  $\lambda_{air}$/nm&  \multicolumn{2}{c}{Present}& Tachiev & Hibbert & Bautista  & Musielok  & Bridges  \\
\cline{4-5}
 &  & & B & C & \& Fischer$^a$ & et al.$^b$ & et al.$^c$ & et al.$^d$ & \& Wiese$^e$\\
 \hline
$\mathrm{2p^2(^3P)3p~^4S_{3/2}^o}$ & $\mathrm{2p^2(^3P)3s~ ^4P_{3/2}}$ & 744.229 & $-0.429$ & $-0.433$ & $-0.403$ & $-0.386$  &                &  $-0.39$ $\pm$ 0.049 &  $-0.41$ $\pm$ 0.035   \\
$\mathrm{2p^2(^3P)3p~ ^4P_{5/2}^o}$ & $\mathrm{2p^2(^3P)3s~ ^4P_{5/2}}$ & 821.633& $0.116$& $0.108$ & $0.138$ & $0.147$   &                &  0.13 $\pm$ 0.053 &                    \\
$\mathrm{2p^2(^3P)3p~ ^2P_{3/2}^o}$  & $\mathrm{2p^2(^3P)3s~ ^2P_{3/2}}$ & 862.923 & $0.070$& $0.068$ & $0.077$ & $0.090$ &  0.057 $\pm$ 0.046 &  0.051 $\pm$ 0.052 &                    \\
$\mathrm{2p^2(^3P)3p~ ^4D_{5/2}^o}$ & $\mathrm{2p^2(^3P)3s~ ^4P_{3/2}}$ & 868.340 & $0.084$& $0.079$& $0.106$ & $0.116$    &  0.14 $\pm$ 0.047  &  0.054 $\pm$ 0.047 &  0.069 $\pm$ 0.035   \\
$\mathrm{2p^2(^3P)3d~ ^4F_{5/2}}$  & $\mathrm{2p^2(^3P)3p~ ^4D_{3/2}^o}$ & 1010.89 & $0.410$ & $0.406$ & $0.444$ & $0.443$ &                &  0.42 $\pm$ 0.056 &  0.53 $\pm$ 0.065      \\
\hline
\end{tabular}
\tablecomments{The table lists: upper and lower configurations along with their spectroscopic terms; wavelengths in air $\lambda_{air}$; weighted oscillator strength log($gf$). $^a$\citet{tachiev.2002}; $^b$\citet{hibbert.1991};
$^c$\citet{2022arXiv220614095B}; $^d$\citet{musielok.1995}; $^e$\citet{bridges.2010}.} 
\end{table*}

\begin{figure}
\centering
\includegraphics[scale=0.315]{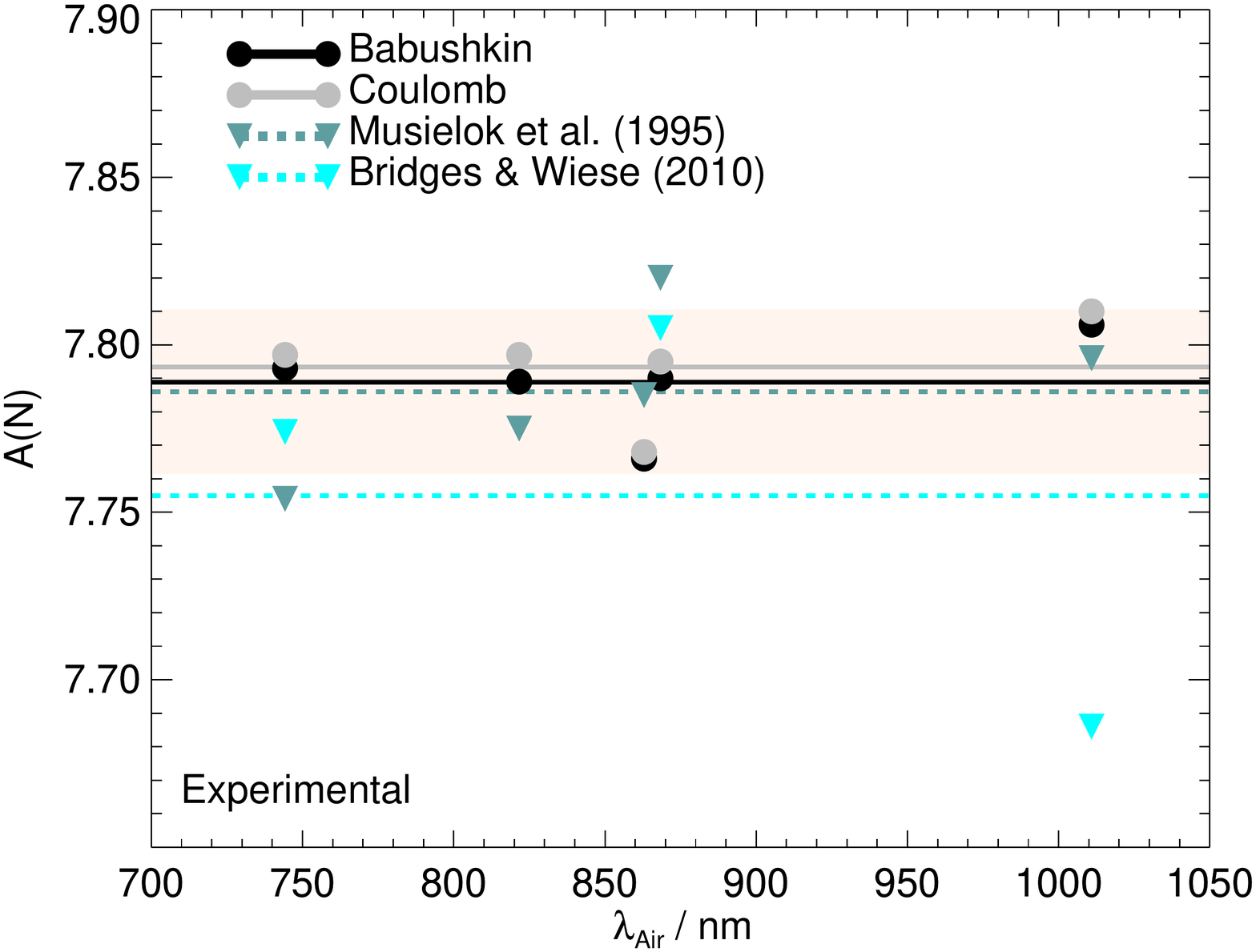}
\includegraphics[scale=0.315]{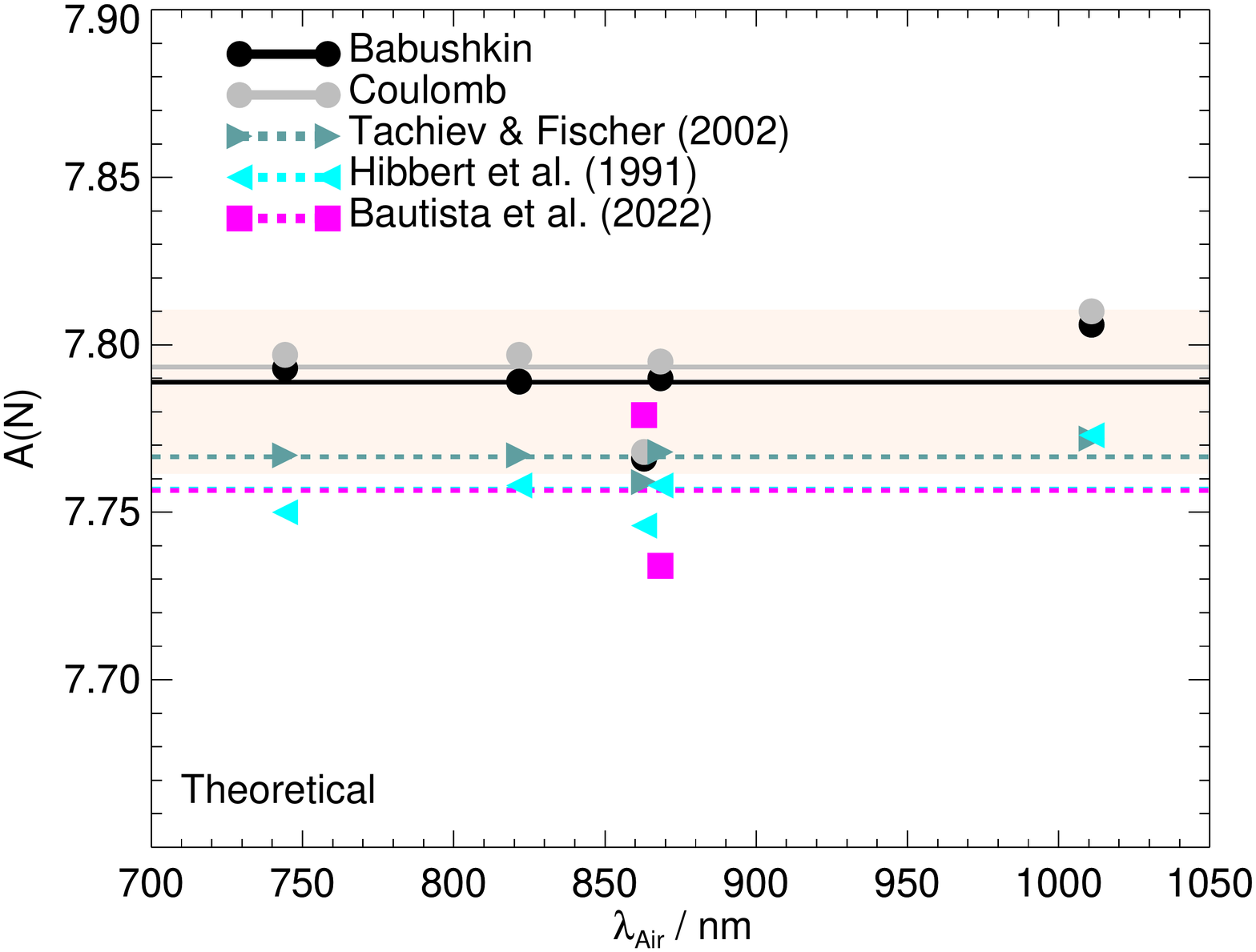}
\caption{Solar nitrogen abundance inferred using different transition probabilities
given in Table~\ref{tab:abundance}. Both panels illustrate results based on the theoretical
transition data computed in the present work. The left panel also includes
results based on the experimental data of \citet{musielok.1995} and \citet{bridges.2010}, 
while the right panel includes
results based on the theoretical data of \citet{tachiev.2002}, \citet{hibbert.1991},
and \citet{2022arXiv220614095B}.
Horizontal lines show the mean abundance inferred from each data set.
In both panels, the shaded region shows the standard deviation of the result based 
on the experimental data of \citet{musielok.1995}, centred on the mean.}
\label{fig:abundance}
\end{figure}

Fig.~\ref{fig:abundance} compares the data in Table~\ref{tab:abundance}
in terms of the solar nitrogen abundance, using Equations~\ref{eq:abcor} and \ref{eq:delta}.
The upper panel compares the results based on the present calculations to the 
those based on the experimental measurements of both \cite{musielok.1995} and \cite{bridges.2010}.
We see that the mean solar nitrogen abundance from the present calculations (in either gauge) agree
well with those from \cite{musielok.1995}. 
Interestingly, this analysis reveals that the transition probability for the 1010.89 nm line
from \cite{bridges.2010} is significantly overestimated, as it would imply a solar nitrogen abundance
significantly lower than what is implied by the other \ion{N}{1} lines.

The lower panel of Fig.~\ref{fig:abundance} compares the results based on the present calculations to the 
those based on the theoretical calculations of \citet{tachiev.2002}, \citet{hibbert.1991},
and \citet{2022arXiv220614095B}. 
We see that the mean solar nitrogen abundance from the present calculations (in either gauge) are
significantly higher than those from the other three data sets.
In particular, this analysis suggests the transition probability for the 868.340 nm line from \cite{2022arXiv220614095B} 
{might be overestimated}, as it would imply a solar nitrogen abundance significantly lower than what is implied by the other \ion{N}{1} lines, for all of {the other theoretical} transition probability data sets.

Our new calculations suggest $\lgeps{}=7.79$ from \ion{N}{1} lines, with the results from both the
Babushkin gauge and Coulomb gauge in agreement to $0.01\,\mathrm{dex}$.
This is an increase of $0.02\,\mathrm{dex}$ 
over the result from \citet{amarsi.2020} based on the transition probabilities given by \citet{tachiev.2002}.
Factoring in also the results from molecular lines
($\lgeps{}=7.89$; \citealt{2021A&A...656A.113A}), our mean solar nitrogen
abundance becomes $7.84$, an increase of $0.01\,\mathrm{dex}$ over the result given
in \citet{asplund.2021}. The rather large difference between the abundances inferred from atomic 
and molecular lines discussed in \citet{2021A&A...656A.113A} are thus probably
not caused by errors in the \ion{N}{1} transition probabilities.

\section{Conclusions}
In the present work, MCDHF and complementary RCI calculations have been performed for the lowest 103 states of \ion{N}{1}; extending the models of earlier state-of-the-art calculations. Excitation energies, lifetimes, wavelengths, line strengths, transition rates, and weighted oscillator strengths have been systematically computed and provided. 

Comparing the excitation energies with experimental data provided by NIST-ASD, the average relative differences of the computed energy levels are 0.07\%. 
The accuracy of the transition data is evaluated based on the relative
differences of the computed transition rates in the Coulomb and Babuskin forms, d$T$, and by extensive comparisons with previous 
theoretical and experimental results.
%\ama{[Add something, somewhere, about testing on the solar spectrum?]}
A statistical analysis of the uncertainties d$T$ of the E1 transitions
is performed and the mean d$T$ for transitions with 
$A > 10^2$ s$^{-1}$ is estimated to be {0.107} ($\sigma = 0.18$), 
and {0.045} ($\sigma$ = 0.08) for transitions with $A > 10^6$ s$^{-1}$.
The agreement of the experimental and present theoretical transition properties, for example, oscillator strengths, transition rates, and line strengths, is overall good. But for some weaker transitions, e.g. intercombination transitions, significant discrepancies are present. Such transitions are subject to strong cancellation effects and cannot be properly considered in the present calculations. An improved methodology is needed to further decrease the uncertainties of the respective transition data.

In addition, the present atomic data were employed in an analysis of the solar nitrogen abundance.
Our new data suggest a mean solar nitrogen abundance $A$(N)= 7.79 from five \ion{N}{1}
lines, with the results from both the Babushkin gauge
and Coulomb gauge in agreement to 0.01 dex.
The new abundance {agrees} well with {that} obtained from
the experimental measurements of \cite{musielok.1995},
while are higher than those from the other theoretical data sets.
However, the large difference between
the abundances inferred from atomic and molecular lines
($A$(N)= 7.89, \cite{2021A&A...656A.113A}) {still
exits} and it is probably not caused by errors in the
\ion{N}{1} transition probabilities.

%At last, the transition parameters of the spectral lines used for solar abundance analysis are provided and compared with values from other calculations.
%\jon{Comment: can you say something conclusive/quantitative, e.g. the 1010.89 line or the general impact of this data set? Maybe not interesting enough, just in case something is worth mentioning here.}
%The extensive and relatively accurate data set on N I reported in this work should be useful for LTE and non-LTE modeling and spectroscopic analysis of astrophysical and laboratory plasma in general. 
%[Comment: I re-wrote this a bit, but not sure if it is meant as a final closing paragraph, or one that is specifically focused on astrophysics?]}

\begin{acknowledgments}
MCL would like to acknowledge the support from the Guangdong Basic and Applied Basic Research Foundation (2022A1515110043) ,the Professorial and Doctoral Scientific Research Foundation of Huizhou University No.158020137.
AMA and JG gratefully acknowledge support from the Swedish Research Council (VR 2020-03940, 2020-05467).{We would also like to thank the anonymous referees for their useful comments that helped improve the original manuscript.}
\end{acknowledgments}

\bibliography{main}{}
\bibliographystyle{aasjournal}
%\input{tab_energy}
%\appendix
%\section{Atomic state function composition in $LS$-coupling, energy levels, and lifetimes for the \ion{N}{1}.}\label{data}
%\renewcommand{\thetable}{A\arabic{table}}
%\input{tab_energy}
%\input{tab_transition}
%\input{tab_com}
%% the collaboration and author truncation commands are used. It has to

\end{document}